\documentclass[final]{siamltex}

\usepackage{dcolumn}
\usepackage{graphicx}
\usepackage{rotating}
\usepackage{multirow}
\usepackage{amsmath,amsfonts,amssymb}
\usepackage{algorithmic}

\newcommand{\e}{\mathrm{e}}
\renewcommand{\d}{\mathrm{d}}
\newcommand{\etal}{\textit{et~al.}}
\newcommand{\defn}{\textit}
\newcommand{\half}{\mbox{$\frac12$}}
\newcommand{\erf}{\mathop\mathrm{erf}}
\newcommand{\erfc}{\mathop\mathrm{erfc}}
\newcommand{\Ord}{\mathrm{O}}
\newcommand{\Expect}[1]{\ensuremath{\mathbf{E}\left[ #1 \right]}}
\newcommand{\xmin}{\ensuremath{x_{\min}}}
\newcommand{\ntail}{\ensuremath{n_{\rm tail}}}
\newcommand{\Like}{\ensuremath{L}}
\newcommand{\Loglike}{\ensuremath{\mathcal{L}}}
\newcommand{\tbf}{\textbf}

\newcommand{\punctspace}{\,}

\title{Power-law distributions in empirical data}
\author{Aaron Clauset\thanks{Santa Fe Institute, 1399 Hyde Park Road, Santa Fe, NM 87501,
 USA and Department of Computer Science, University of New Mexico, Albuquerque, NM 87131, USA}
 \and Cosma Rohilla Shalizi\thanks{Department of Statistics, Carnegie Mellon University, Pittsburgh, PA 15213, USA}
 \and M. E. J. Newman\thanks{Department of Physics and Center for the Study of Complex Systems, University of Michigan, Ann Arbor, MI 48109, USA}
 }

\begin{document}

\maketitle

\begin{abstract}
 Power-law distributions occur in many situations of scientific interest
 and have significant consequences for our understanding of natural and
 man-made phenomena.  Unfortunately, the detection and characterization of
 power laws is complicated by the large fluctuations that occur in the
 tail of the distribution---the part of the distribution representing
 large but rare events---and by the difficulty of identifying the range
 over which power-law behavior holds.  Commonly used methods for analyzing
 power-law data, such as least-squares fitting, can produce substantially
 inaccurate estimates of parameters for power-law distributions, and even
 in cases where such methods return accurate answers they are still
 unsatisfactory because they give no indication of whether the data obey a
 power law at all.  Here we present a principled statistical framework for
 discerning and quantifying power-law behavior in empirical data.  Our
 approach combines maximum-likelihood fitting methods with goodness-of-fit
 tests based on the Kolmogorov-Smirnov statistic and likelihood ratios.
 We evaluate the effectiveness of the approach with tests on synthetic
 data and give critical comparisons to previous approaches.  We also apply
 the proposed methods to twenty-four real-world data sets from a range of
 different disciplines, each of which has been conjectured to follow a
 power-law distribution.  In some cases we find these conjectures to be
 consistent with the data while in others the power law is ruled out.
\end{abstract}

\begin{keywords}
 Power-law distributions; Pareto; Zipf; maximum likelihood; heavy-tailed
 distributions; likelihood ratio test; model selection
\end{keywords}

\begin{AMS} 62-07, 62P99, 65C05, 62F99 \end{AMS}

\pagestyle{myheadings}
\thispagestyle{plain}
\markboth{A. Clauset, C. R. Shalizi and M. E. J. Newman}{Power-law distributions in empirical data}

\section{Introduction}

Many empirical quantities cluster around a typical value.  The speeds of
cars on a highway, the weights of apples in a store, air pressure, sea
level, the temperature in New York at noon on Midsummer's Day.  All of
these things vary somewhat, but their distributions place a negligible
amount of probability far from the typical value, making the typical value
representative of most observations.  For instance, it is a useful
statement to say that an adult male American is about 180cm tall because no
one deviates very far from this size.  Even the largest deviations, which
are exceptionally rare, are still only about a factor of two from the mean
in either direction and hence the distribution can be well-characterized by
quoting just its mean and standard deviation.

Not all distributions fit this pattern, however, and while those that do
not are often considered problematic or defective for just that reason,
they are at the same time some of the most interesting of all scientific
observations.  The fact that they cannot be characterized as simply as
other measurements is often a sign of complex underlying processes that
merit further study.

Among such distributions, the \defn{power law} has attracted particular
attention over the years for its mathematical properties, which sometimes
lead to surprising physical consequences, and for its appearance in a
diverse range of natural and man-made phenomena.  The populations of
cities, the intensities of earthquakes, and the sizes of power outages, for
example, are all thought to have power-law distributions.  Quantities such
as these are not well characterized by their typical or average values.
For instance, according to the 2000 US Census, the average population of a
city, town, or village in the United States is 8226.  But this statement is
not a useful one for most purposes because a significant fraction of the
total population lives in cities (New York, Los Angeles, etc.)\ whose
population is larger by several orders of magnitude.  Extensive discussions
of this and other properties of power laws can be found in the reviews by
Mitzenmacher~\cite{Mitzenmacher04}, Newman~\cite{MEJN-on-power-laws}, and
Sornette~\cite{Sornette06pl}, and references therein.

Mathematically, a quantity $x$ obeys a power law if it is drawn from a
probability distribution
\begin{equation}
p(x) \propto x^{-\alpha},
\label{eq:powerlaw}
\end{equation}
where $\alpha$ is a constant parameter of the distribution known as the
\defn{exponent} or \defn{scaling parameter}.  The scaling parameter
typically lies in the range $2<\alpha<3$, although there are occasional
exceptions.

In practice, few empirical phenomena obey power laws for all values of~$x$.
More often the power law applies only for values greater than some
minimum~$\xmin$.  In such cases we say that the \defn{tail} of the
distribution follows a power law.

In this article, we address a recurring issue in the scientific literature,
the question of how to recognize a power law when we see one.  In practice,
we can rarely, if ever, be certain that an observed quantity is drawn from
a power-law distribution.  The most we can say is that our observations are
consistent with the hypothesis that $x$ is drawn from a distribution of the
form of Eq.~\eqref{eq:powerlaw}.  In some cases we may also be able to rule
out some other competing hypotheses.  In this paper we describe in detail a
set of statistical techniques that allow one to reach conclusions like
these, as well as methods for calculating the parameters of power laws when
we find them.  Many of the methods we describe have been discussed
previously; our goal here is to bring them together to create a complete
procedure for the analysis of power-law data.  A short description
summarizing this procedure is given in Box~1.  Software implementing it is
also available on-line.\footnote{See
 \texttt{http://www.santafe.edu/\~{}aaronc/powerlaws/}.}

Practicing what we preach, we also apply our methods to a large number of
data sets describing observations of real-world phenomena that have at one
time or another been claimed to follow power laws.  In the process, we
demonstrate that several of them cannot reasonably be considered to follow
power laws, while for others the power-law hypothesis appears to be a good
one, or at least is not firmly ruled out.

\newsavebox{\recipe}
\sbox{\recipe}{\fbox{%
\begin{minipage}{12.6cm}
\setlength{\parskip}{4pt}
\begin{center}
\textbf{Box 1: Recipe for analyzing power-law distributed data}
\end{center}

\small
This paper contains much technical detail.  In broad outline, however, the
recipe we propose for the analysis of power-law data is straightforward and
goes as follows.

\begin{enumerate}
\item Estimate the parameters $\xmin$ and $\alpha$ of the power-law model
 using the methods described in Section~\ref{sec:fitting}.
\item Calculate the goodness-of-fit between the data and the power law
 using the method described in Section~\ref{sec:hypothesis}.  If the
 resulting $p$-value is greater than $0.1$ the power law is a plausible
 hypothesis for the data, otherwise it is rejected.
\item Compare the power law with alternative hypotheses via a likelihood
 ratio test, as described in Section~\ref{sec:alttesting}.  For each
 alternative, if the calculated likelihood ratio is significantly
 different from zero, then its sign indicates whether the alternative is
 favored over the power-law model or not.
\end{enumerate}

Step~3, the likelihood ratio test for alternative hypotheses, could in
principle be replaced with any of several other established and
statistically principled approaches for model comparison, such as a fully
Bayesian approach~\cite{kass:raftery:1995}, a cross-validation
approach~\cite{stone:1974}, or a minimum description length
approach~\cite{grunwald:2007}, although none of these methods are described
here.
\end{minipage}
}}
\begin{table*}
\usebox{\recipe}
\end{table*}

\section{Definitions}
\label{sec:definitions}
We begin our discussion of the analysis of power-law distributed data with
some brief definitions of the basic quantities involved.

Power-law distributions come in two basic flavors: continuous distributions
governing continuous real numbers and discrete distributions where the
quantity of interest can take only a discrete set of values, typically
positive integers.

Let $x$ represent the quantity whose distribution we are interested in.  A
continuous power-law distribution is one described by a probability
density~$p(x)$ such that
\begin{equation}
 p(x)\>\d x = \Pr(x \le X < x+\d x)
            = Cx^{-\alpha}\>\d x\punctspace,
\label{eq:contpl}
\end{equation}
where $X$ is the observed value and $C$ is a normalization constant.
Clearly this density diverges as $x\to0$ so Eq.~\eqref{eq:contpl} cannot
hold for all $x\ge0$; there must be some lower bound to the power-law
behavior.  We will denote this bound by $\xmin$.  Then, provided
$\alpha>1$, it is straightforward to calculate the normalizing constant and
we find that
\begin{equation}
\label{eq:full:continuous}
p(x) = \frac{\alpha-1}{\xmin}\,
 \left( \frac{x}{\xmin} \right)^{-\alpha}\punctspace.
\end{equation}

In the discrete case, $x$~can take only a discrete set of values.  In this
paper we consider only the case of integer values with a probability
distribution of the form
\begin{equation}
p(x) = \Pr(X=x) = Cx^{-\alpha}\punctspace.
\end{equation}
Again this distribution diverges at zero, so there must be a lower
bound~$\xmin>0$ on the power-law behavior.  Calculating the normalizing
constant, we then find that
\begin{equation}
\label{eq:full:discrete}
p(x) = {x^{-\alpha}\over\zeta(\alpha,\xmin)}\punctspace,
\end{equation}
where
\begin{equation}
\zeta(\alpha,\xmin) = \sum_{n=0}^\infty (n+\xmin)^{-\alpha}
\end{equation}
is the generalized or Hurwitz zeta function.
Table~\ref{table:distributions} summarizes the basic functional forms and
normalization constants for these and several other distributions that will
be useful.

In many cases it is useful to consider also the complementary cumulative
distribution function or CDF of a power-law distributed variable, which we
denote $P(x)$ and which for both continuous and discrete cases is defined
to be $P(x) = \Pr(X\ge x)$.  For instance, in the continuous case
\begin{equation}
\label{eq:cdf:continuous}
P(x) = \int_x^\infty p(x') \>\d x'
    = \biggl( {x\over x_\mathrm{min}} \biggr)^{-\alpha+1}\punctspace.
\end{equation}
In the discrete case
\begin{equation}
\label{eq:cdf:discrete}
P(x) = {\zeta(\alpha,x)\over\zeta(\alpha,\xmin)}\punctspace.
\end{equation}

Because formulas for continuous distributions, such as
Eq.~\eqref{eq:full:continuous}, tend to be simpler than those for discrete
distributions, it is common to approximate discrete power-law behavior with
its continuous counterpart for the sake of mathematical convenience.  But a
word of caution is in order: there are several different ways to
approximate a discrete power law by a continuous one and though some of
them give reasonable results, others do not.  One relatively reliable
method is to treat an integer power law as if the values of~$x$ were
generated from a continuous power law then rounded to the nearest integer.
This approach gives quite accurate results in many applications.  Other
approximations, however, such as truncating (rounding down), or simply
assuming that the probabilities of generation of integer values in the
discrete and continuous cases are proportional, give poor results and
should be avoided.

Where appropriate we will discuss the use of continuous approximations for
the discrete power law in the sections that follow, particularly in
Section~\ref{sec:fitting} on the estimation of best-fit values for the
scaling parameter from observational data and in
Appendix~\ref{appendix:deviates} on the generation of power-law distributed
random numbers.

\setlength{\tabcolsep}{4pt}
\begin{table*}
\begin{center}
\begin{tabular}{l|l|cc}
& \multirow{2}{*}{name} & \multicolumn{2}{c}{distribution $p(x)=Cf(x)$} \\
&                       & $f(x)$ & $C$                                   \\
\hline
\multirow{4}{*}{\rotatebox{90}{continuous\hspace{3.2em}}}
 & \multirow{2}{5em}{power law}
 & \multirow{2}{*}{$x^{-\alpha}$}
 & \multirow{2}{*}{$(\alpha-1)\xmin^{\alpha-1}$} \\
& & & \\

 & \multirow{2}{5em}{power law with cutoff}
 & \multirow{2}{*}{$x^{-\alpha}\e^{-\lambda x}$}
 & \multirow{2}{*}{${\lambda^{1-\alpha}\over\Gamma(1-\alpha,\lambda \xmin)}$} \\
& & & \\

 & \multirow{2}{5em}{exponential}
 & \multirow{2}{*}{$\e^{-\lambda x}$}
 & \multirow{2}{*}{$\lambda \e^{\lambda \xmin}$} \\
& & & \\

 & \multirow{2}{5em}{stretched exponential}
 & \multirow{2}{*}{$x^{\beta-1} \e^{-\lambda x^\beta}$}
 & \multirow{2}{*}{$\beta\lambda \e^{\lambda \xmin^\beta}$} \\
& & & \\

 & \multirow{2}{5em}{log-normal}
 & \multirow{2}{*}{${1\over x} \exp\left[ -{(\ln x-\mu)^2\over2\sigma^2} \right]$}
 & \multirow{2}{*}{$\sqrt{{2\over\pi\sigma^2}}\Bigl[ \erfc\left( {\ln x_{\mathrm{min}}-\mu\over\sqrt{2}\sigma} \right) \Bigr]^{-1}$} \\
& & & \\

\hline
\multirow{4}{*}{\rotatebox{90}{discrete\hspace{3em}}}
 & \multirow{2}{5em}{power law}
 & \multirow{2}{*}{$x^{-\alpha}$}
 & \multirow{2}{*}{$1/\zeta(\alpha,\xmin)$} \\
& & & \\

 & \multirow{2}{5em}{Yule distribution}
 & \multirow{2}{*}{${\Gamma(x)\over\Gamma(x+\alpha)}$}
 & \multirow{2}{*}{$(\alpha-1){\Gamma(\xmin+\alpha-1)\over\Gamma(\xmin)}$} \\
& & & \\

 & \multirow{2}{5em}{exponential}
 & \multirow{2}{*}{$\e^{-\lambda x}$}
 & \multirow{2}{*}{$(1-\e^{-\lambda})\,\e^{\lambda \xmin}$} \\
& & & \\

 & \multirow{2}{5em}{Poisson}
 & \multirow{2}{*}{$\mu^x/x!$}
 & \multirow{2}{*}{${\left[\e^{\mu} - \sum_{k=0}^{\xmin-1}{\frac{\mu^k}{k!}}\right]}^{-1}$} \\
& & & \\
\end{tabular}
\end{center}
\caption{Definition of the power-law distribution and several other common
 statistical distributions.  For each distribution we give the
 basic functional form~$f(x)$ and the appropriate normalization
 constant~$C$ such that $\int_{\xmin}^\infty Cf(x)\>\d x=1$ for the
 continuous case or $\sum_{x=\xmin}^\infty Cf(x) = 1$ for the discrete
 case.}
\label{table:distributions}
\end{table*}

\section{Fitting power laws to empirical data}
\label{sec:fitting}
We turn now to the first of the main goals of this paper, the correct fitting
of power-law forms to empirical distributions.  Studies of empirical
distributions that follow power laws usually give some estimate of the scaling
parameter~$\alpha$ and occasionally also of the lower-bound on the scaling
region~$\xmin$.  The tool most often used for this task is the simple
histogram.  Taking the logarithm of both sides of Eq.~\eqref{eq:powerlaw}, we
see that the power-law distribution obeys $\ln p(x) = \alpha \ln x +
\mbox{constant}$, implying that it follows a straight line on a doubly
logarithmic plot.  A common way to probe for power-law behavior, therefore, is
to measure the quantity of interest~$x$, construct a histogram representing its
frequency distribution, and plot that histogram on doubly logarithmic axes. If
in so doing one discovers a distribution that approximately falls on a straight
line, then one can, if one is feeling particularly bold, assert that the
distribution follows a power law, with a scaling parameter~$\alpha$ given by
the absolute slope of the straight line.  Typically this slope is extracted by
performing a least-squares linear regression on the logarithm of the histogram.
This procedure dates back to Pareto's work on the distribution of wealth at the
close of the 19th century~\cite{Arnold-on-Pareto-distributions}.

Unfortunately, this method and other variations on the same theme generate
significant systematic errors under relatively common conditions, as
discussed in Appendix~\ref{appendix:linear_regression}, and as a
consequence the results they give cannot not be trusted.  In this section
we describe a generally accurate method for estimating the parameters of a
power-law distribution.  In Section~\ref{sec:hypothesis} we study the
equally important question of how to determine whether a given data set
really does follow a power law at all.

\subsection{Estimating the scaling parameter}
\label{sec:estalpha}
First, let us consider the estimation of the scaling parameter~$\alpha$.
Estimating~$\alpha$ correctly requires, as we will see, a value for the
lower bound~$\xmin$ of power-law behavior in the data.  For the moment, let
us assume that this value is known.  In cases where it is unknown, we can
estimate it from the data as well, and we will consider methods for doing
this in Section~\ref{sec:xmin}.

The method of choice for fitting parametrized models such as power-law
distributions to observed data is the \defn{method of maximum likelihood},
which provably gives accurate parameter estimates in the limit of large
sample size~\cite{Wasserman-all-of-stats,%
 Barndorff-Nielsen-and-Cox-inference-and-asymptotics}.  Assuming that our
data are drawn from a distribution that follows a power law exactly
for~$x\ge \xmin$, we can derive maximum likelihood estimators (MLEs) of the
scaling parameter for both the discrete and continuous cases.  Details of
the derivations are given in Appendix~\ref{appendix:pl_mle}; here our focus
is on their use.

The MLE for the continuous case is~\cite{Muniruzzaman-on-Pareto}
\begin{equation}
\hat{\alpha} = 1 + n \Biggl[ \sum_{i=1}^{n}\ln \frac{x_{i}}{\xmin}
                    \Biggr]^{-1}\punctspace,
\label{eq:contmle}
\end{equation}
where $x_i$, $i=1\ldots n$ are the observed values of~$x$ such that
$x_{i}\ge \xmin$.  Here and elsewhere we use ``hatted'' symbols such
as~$\hat{\alpha}$ to denote estimates derived from data; hatless symbols
denote the true values, which are often unknown in practice.

Equation~\eqref{eq:contmle} is equivalent to the well-known Hill
estimator~\cite{Hill75}, which is known to be asymptotically
normal~\cite{Hall82} and consistent~\cite{Mason82}
(i.e.,~$\hat{\alpha}\to\alpha$ in the limit of large~$n$).  The standard
error on~$\hat{\alpha}$, which is derived from the width of the likelihood
maximum, is
\begin{equation}
\label{eq:mle:sigma}
\sigma = {\hat{\alpha}-1\over\sqrt{n}} + \Ord(1/n)\punctspace,
\end{equation}
where the higher-order correction is positive; see Appendix
\ref{appendix:pl_mle} of this paper or any of
Refs.~\cite{Muniruzzaman-on-Pareto}, \cite{MEJN-on-power-laws},
or~\cite{Wheatland04}.

(We assume in these calculations that $\alpha>1$, since distributions with
$\alpha\le1$ are not normalizable and hence cannot occur in nature.  It is
possible for a probability distribution to go as $x^{-\alpha}$ with
$\alpha\le1$ if the range of~$x$ is bounded above by some cutoff, but different
maximum likelihood estimators are needed to fit such a distribution.)

The MLE for the case where $x$ is a discrete integer variable is less
straightforward.  Ref.~\cite{Seal-1952} and more
recently~\cite{Goldstein-Morris-Yen} treated the special case $\xmin=1$,
showing that the appropriate estimator for~$\alpha$ is given by the
solution to the transcendental equation
\begin{equation}
\label{eq:discrete:mle}
\frac{\zeta'(\hat{\alpha})}{\zeta(\hat{\alpha})}
 = -\frac{1}{n}\sum_{i=1}^{n}\ln x_{i} \punctspace.
\end{equation}
When $\xmin>1$, a similar equation holds, but with the zeta functions replaced
by generalized zetas~\cite{Arnold-on-Pareto-distributions,Bauke07,Clauset07}:
\begin{equation}
\frac{\zeta'(\hat{\alpha},\xmin)}{\zeta(\hat{\alpha},\xmin)} =
-\frac{1}{n}\sum_{i=1}^{n}\ln x_{i}\punctspace,
\label{eq:discmle}
\end{equation}
where the prime denotes differentiation with respect to the first argument.  In
practice, evaluation of $\hat{\alpha}$ requires us to solve this equation
numerically.  Alternatively, one can estimate $\alpha$ by direct numerical
maximization of the likelihood function itself, or equivalently of its
logarithm (which is usually simpler):
\begin{align}
\Loglike(\alpha) = -n\ln\zeta(\alpha,\xmin) -
\alpha\sum_{i=1}^{n} \ln x_{i}\punctspace.
\label{eq:discll}
\end{align}
To find an estimate for the standard error on~$\hat{\alpha}$ in the
discrete case, we make a quadratic approximation to the log-likelihood at
its maximum and take the standard deviation of the resulting Gaussian form
for the likelihood as our error estimate (an approach justified by general
theorems on the large-sample-size behavior of maximum likelihood
estimates---see, for example, Theorem~\ref{theorem:asymptotic-dist-of-mle}
of Appendix~\ref{appendix:pl_mle}).  The result is
\begin{equation}
\sigma = {1\over\sqrt{ n\biggl[ \displaystyle
                  \frac{\zeta''(\hat{\alpha},\xmin)}%
                 {\zeta(\hat{\alpha},\xmin)} - 
                 \biggl( \frac{\zeta'(\hat{\alpha},\xmin)}%
	         {\zeta(\hat{\alpha},\xmin)} \biggr)^2 \biggl] }}\punctspace,
\label{eq:mle:d:sigma}
\end{equation}
which is straightforward to evaluate once we have~$\hat{\alpha}$.
Alternatively, Eq.~\eqref{eq:mle:sigma} yields roughly similar results for
reasonably large~$n$ and~$\xmin$.

Although there is no exact closed-form expression for~$\hat{\alpha}$ in the
discrete case, an approximate expression can be derived using the approach
mentioned in Section~\ref{sec:definitions} in which true power-law
distributed integers are approximated as continuous reals rounded to the
nearest integer.  The details of the derivation are given in
Appendix~\ref{appendix:pl_mle}.  The result is
\begin{equation}
\hat{\alpha} \simeq 1 + n \Biggl[ \sum_{i=1}^n \ln {x_i\over \xmin-\half}
             \Biggr]^{-1}\punctspace.
\label{eq:approxmle}
\end{equation}
This expression is considerably easier to evaluate than the exact discrete
MLE and can be useful in cases where high accuracy is not needed.  The size
of the bias introduced by the approximation is discussed in
Appendix~\ref{appendix:pl_mle}.  In practice this estimator gives quite
good results; in our own experiments we have found it to give results
accurate to about 1\% or better provided $\xmin\gtrsim6$.  An estimate of
the statistical error on $\hat{\alpha}$ (which is quite separate from the
systematic error introduced by the approximation) can be calculated by
employing Eq.~\eqref{eq:mle:sigma} again.

Another approach taken by some authors is simply to pretend that discrete
data are in fact continuous and then use the MLE for continuous data,
Eq.~\eqref{eq:contmle}, to calculate~$\hat{\alpha}$.  This approach,
however, gives significantly less accurate values of $\hat{\alpha}$ than
Eq.~\eqref{eq:approxmle} and, given that it is no easier to implement, we
see no reason to use it in any circumstances.\footnote{The error involved
 can be shown to decay as $\Ord\bigl(\xmin^{-1}\bigr)$, while the error on
 Eq.~\eqref{eq:approxmle} decays much faster, as
 $\Ord\bigl(\xmin^{-2}\bigr)$.  In our own experiments we have found that
 for typical values of~$\alpha$ we need $\xmin\gtrsim100$ before
 Eq.~\eqref{eq:contmle} becomes accurate to about 1\%, as compared to
 $\xmin\gtrsim6$ for Eq.~\eqref{eq:approxmle}.}

\subsection{Performance of scaling parameter estimators}

To demonstrate the working of the estimators described above, we now test
their ability to extract the known scaling parameters of synthetic
power-law data.  Note that in practical situations we usually do not know
\textit{a priori}, as we do in the calculations of this section, that our
data are power-law distributed.  In that case, our MLEs will give us no
warning that our fits are wrong: they tell us only the best fit to the
power-law form, not whether the power law is in fact a good model for the
data.  Other methods are needed to address the latter question, which are
discussed in Sections~\ref{sec:hypothesis} and~\ref{sec:alttesting}.

\begin{figure}[t]
\begin{center}
\includegraphics[scale=0.6]{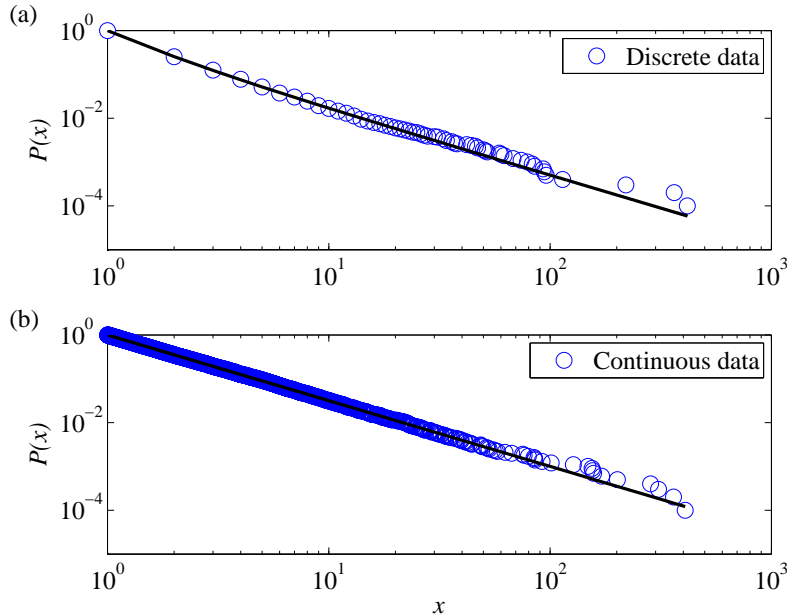}
\end{center}
\caption{Points represent the cumulative density functions $P(x)$ for
 synthetic data sets distributed according to (a)~a discrete power law and
 (b)~a continuous power law, both with $\alpha=2.5$ and $\xmin=1$.  Solid
 lines represent best fits to the data using the methods described in the
 text.}
\label{fig:estimates}
\end{figure}

\begin{table}
\centering
\begin{tabular}{lc|ccccc}
                            & & & est. $\alpha$ & & est. $\alpha$ \\
             method & notes & & (discrete) & & (continuous) \\
\hline
LS + PDF & const.\ width & & 1.5(1) & & 1.39(5) \\
LS + CDF & const.\ width & & 2.37(2) & & 2.480(4) \\
LS + PDF & log.\ width & & 1.5(1) & & 1.19(2) \\
LS + CDF & rank-freq. & & 2.570(6) & & 2.4869(3) \\
cont. MLE & -- & & 4.46(3) & & {\bf 2.50(2)} \\
disc. MLE & -- & & {\bf 2.49(2)} & & 2.19(1)
\end{tabular}
\caption{Estimates of the scaling parameter~$\alpha$ using various
 estimators for discrete and continuous synthetic data with $\alpha=2.5$,
 $\xmin=1$, and $n=10\,000$ data points.  LS denotes a least-squares
 fit to the logarithm of the probability.  For the continuous data, the
 probability density function (PDF) was computed in two different ways,
 using bins of constant width $0.1$ and using up to 500 bins of
 exponentially increasing width (so-called ``logarithmic binning'').  The
 cumulative distribution function (CDF) was also
 calculated in two ways, as the cumulation of the fixed-width histogram
 and as a standard rank-frequency function.  In applying the discrete
 MLE to the continuous data, the non-integer part of each measurement was
 discarded.  Accurate estimates are shown in \tbf{bold}.}
\label{table:estimates}
\end{table}

Using methods described in Appendix~\ref{appendix:deviates}, we have
generated two sets of power-law distributed data, one continuous and one
discrete, with~$\alpha=2.5$, $\xmin=1$ and $n=10\,000$ in each case.
Applying our MLEs to these data we calculate that \mbox{$\hat{\alpha} =
 2.50(2)$} for the continuous case and $\hat{\alpha} = 2.49(2)$ for the
discrete case.  (Values in parentheses indicate the uncertainty in the
final digit, calculated from Eqs.~\eqref{eq:mle:sigma}
and~\eqref{eq:mle:d:sigma}.)  These estimates agree well with the known
true scaling parameter from which the data were generated.
Figure~\ref{fig:estimates} shows the distributions of the two data sets
along with fits using the estimated parameters.  (In this and all
subsequent such plots, we show not the probability density function but the
complementary cumulative density function~$P(x)$.  Generally, the visual
form of the CDF is more robust than that of the PDF against fluctuations
due to finite sample sizes, particularly in the tail of the distribution.)

\begin{figure}[t]
\begin{center}
\includegraphics[scale=0.6]{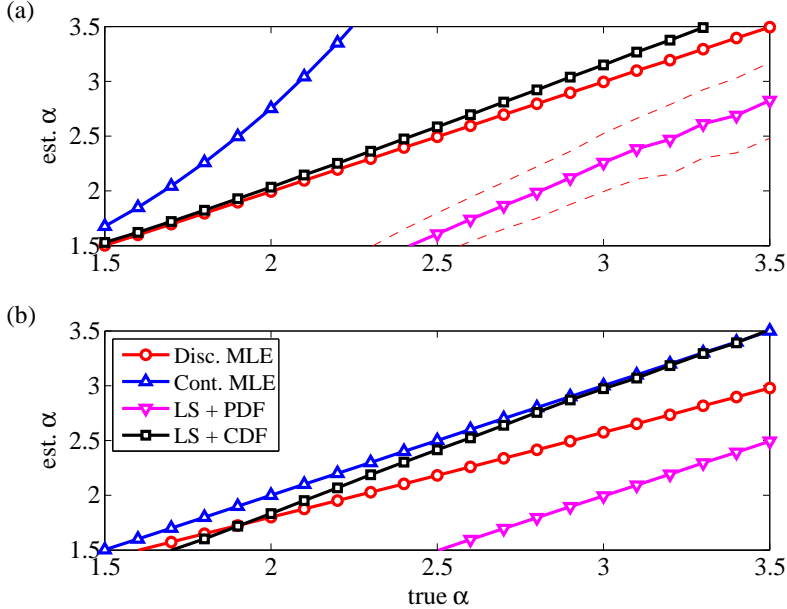}
\end{center}
\caption{Values of the scaling parameter estimated using four of the
 methods of Table~\ref{table:estimates} (we omit the methods based on
 logarithmic bins for the PDF and constant width bins for the CDF) for
 $n=10\,000$ observations drawn from (a)~discrete and (b)~continuous
 power-law distributions with $\xmin=1$.  We omit error bars where they
 are smaller than the symbol size.  Clearly, only the discrete MLE is
 accurate for discrete data, and the continuous MLE for continuous data.}
\label{fig:bias}
\end{figure}

In Table~\ref{table:estimates} we compare the results given by the MLEs to
estimates of the scaling parameter made using several alternative methods
based on linear regression: a straight-line fit to the slope of a
log-transformed histogram, a fit to the slope of a histogram with
``logarithmic bins'' (bins whose width increases in proportion to~$x$,
thereby reducing fluctuations in the tail of the histogram), a fit to the
slope of the CDF calculated with constant width bins, and a fit to the
slope of the CDF calculated without any bins (also called a
``rank-frequency plot''---see~\cite{MEJN-on-power-laws}).  As the table
shows, the MLEs give the best results while the regression methods all give
significantly biased values, except perhaps for the fits to the CDF, which
produce biased estimates in the discrete case but do reasonably well in the
continuous case.  Moreover, in each case where the estimate is biased, the
corresponding error estimate gives no warning of the bias: there is nothing
to alert unwary experimenters to the fact that their results are
substantially incorrect.  Figure~\ref{fig:bias} extends these results
graphically by showing how the estimators fare as a function of the true
$\alpha$ for a large selection of synthetic data sets with $n=10\,000$
observations each.


Finally, we note that the maximum likelihood estimators are only guaranteed
to be unbiased in the asymptotic limit of large sample size, $n\to\infty$.
For finite data sets, biases are present but decay as $\Ord(n^{-1})$ for
any choice of~$\xmin$ (see Appendix~\ref{appendix:pl_mle} and
Fig.~\ref{fig:nconverge}).  For very small data sets, such biases can be
significant but in most practical situations they can be ignored because
they are much smaller than the statistical error of the estimator, which
decays as $\Ord(n^{-1/2})$.  Our experience suggests that $n\gtrsim50$ is a
reasonable rule of thumb for extracting reliable parameter estimates.  For
the examples shown in Fig.~\ref{fig:nconverge} this gives estimates of
$\alpha$ accurate to about 1\%.  Data sets smaller than this should be
treated with caution.  Note, however, that there are more important reasons
to treat small data sets with caution.  Namely, it is difficult to rule out
alternative fits to such data, even when they are truly power-law
distributed, and conversely the power-law form may appear to be a good fit
even when the data are drawn from a non-power-law distribution.  We address
these issues in Sections~\ref{sec:hypothesis} and~\ref{sec:alttesting}.

\subsection{Estimating the lower bound on power-law behavior}
\label{sec:xmin}
As we have said it is normally the case that empirical data, if they follow
a power-law distribution at all, do so only for values of $x$ above some
lower bound~$\xmin$.  Before calculating our estimate of the scaling
parameter~$\alpha$, therefore, we need first to discard all samples below
this point so that we are left with only those for which the power-law
model is a valid one.  Thus, if we wish our estimate of~$\alpha$ to be
accurate we will also need an accurate method for estimating~$\xmin$.  If
we choose too low a value for $x_{\min}$ we will get a biased estimate of
the scaling parameter since we will be attempting to fit a power-law model
to non-power-law data.  On the other hand, if we choose too high a value
for $x_{\min}$ we are effectively throwing away legitimate data points
$x_i<\hat{x}_{\min}$, which increases both the statistical error on the
scaling parameter and the bias from finite size effects.

\begin{figure}[t]
\begin{center}
\includegraphics[scale=0.6]{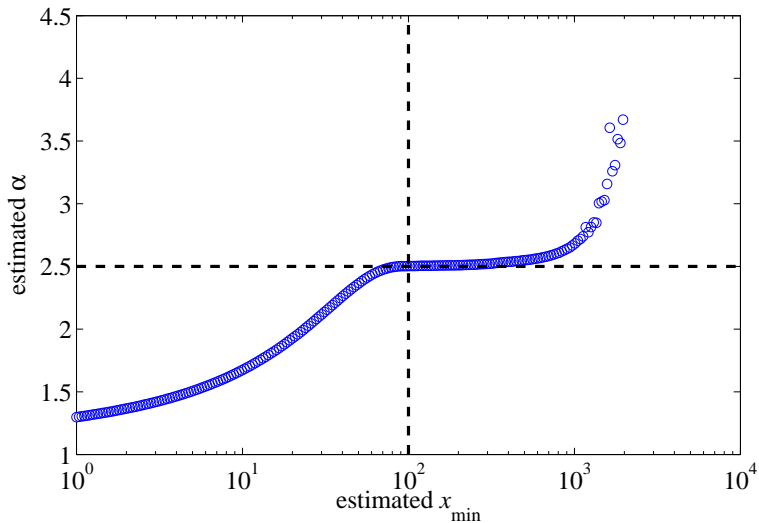}
\end{center}
\caption{Mean of the maximum likelihood estimate for the scaling parameter
 for 5000 samples drawn from the test distribution, Eq.~\eqref{eq:kstest},
 with $\alpha=2.5$, $x_{\min}=100$, and $n=2500$, plotted as a function of
 the value assumed for~$x_{\min}$.  Statistical errors are smaller than
 the data points in all cases.}
\label{fig:xmin_alpha}
\end{figure}

The importance of using the correct value for~$x_{\min}$ is demonstrated in
Fig.~\ref{fig:xmin_alpha}, which shows the maximum likelihood
value~$\hat{\alpha}$ of the scaling parameter averaged over 5000 data sets
of $n=2500$ samples each drawn from the continuous form of
Eq.~\eqref{eq:kstest} with $\alpha=2.5$, as a function of the assumed value
of~$x_{\min}$, where the true value is~100.  As the figure shows, the MLE
gives accurate answers when $x_{\min}$ is chosen exactly equal to the true
value, but deviates rapidly below this point (because the distribution
deviates from power-law) and more slowly above (because of dwindling sample
size).  It would probably be acceptable in this case for $x_{\min}$ to err
a little on the high side (though not too much), but estimates that are too
low could have severe consequences.

The most common ways of choosing $\hat{x}_{\min}$ are either to estimate
visually the point beyond which the PDF or CDF of the distribution becomes
roughly straight on a log-log plot, or to plot $\hat{\alpha}$ (or a related
quantity) as a function of $\hat{x}_{\min}$ and identify a point beyond
which the value appears relatively stable.  But these approaches are
clearly subjective and can be sensitive to noise or fluctuations in the
tail of the distribution---see~\cite{Stoev-Michailidis-Taqqu-max-self-sim}
and references therein.  A more objective and principled approach is
desirable.  Here we review two such methods, one that is specific to
discrete data and is based on a so-called marginal likelihood, and one that
works for either discrete or continuous data and is based on minimizing the
``distance'' between the power-law model and the empirical data.

The first approach, put forward by Handcock and
Jones~\cite{Handcock-Jones-likelihood-of-sexual-networks}, uses a
generalized model to represent all of the observed data, both above and
below~$\hat{x}_{\min}$.  Above~$\hat{x}_{\min}$ the data are modeled by the
standard discrete power-law distribution of Eq.~\eqref{eq:full:discrete};
below~$\hat{x}_{\min}$ each of the $\hat{x}_{\min}-1$ discrete values of
$x$ are modeled by a separate probability $p_k=\Pr(X=k)$ for \mbox{$1\leq
 k<\hat{x}_{\min}$} (or whatever range is appropriate for the problem in
hand).  The MLE for $p_{k}$ is simply the fraction of observations with
value~$k$.  The task then is to find the value for $\hat{x}_{\min}$ such
that this model best fits the observed data.  One cannot, however, fit such
a model to the data directly within the maximum likelihood framework
because the number of model parameters is not fixed: it is equal
to~$\xmin$.\footnote{There is one parameter for each of the $p_k$ plus the
 scaling parameter of the power law.  The normalization constant does not
 count as a parameter, because it is fixed once the values of the other
 parameters are chosen, and~$\xmin$ does not count as a parameter because
 we know its value automatically once we are given a list of the other
 parameters---it is just the length of that list.}  In this kind of
situation, one can always achieve a higher likelihood by increasing the
number of parameters, thus making the model more flexible, so the maximum
likelihood would always be achieved for~$\xmin\to\infty$.  A standard
(Bayesian) approach in such cases is instead to maximize the \defn{marginal
 likelihood} (also called the
\emph{evidence})~\cite{jeffreys:1935,mackay:2003}, i.e.,~the likelihood of
the data given the number of model parameters, integrated over the
parameters' possible values.  Unfortunately, the integral cannot usually be
performed analytically, but one can employ a Laplace or steepest-descent
approximation in which the log-likelihood is expanded to leading
(i.e.,~quadratic) order about its maximum and the resulting Gaussian
integral carried out to yield an expression in terms of the value at the
maximum and the determinant of the appropriate Hessian
matrix~\cite{Tierney-Kass-Kadane}.  Schwarz~\cite{Schwarz-BIC} showed that
the terms involving the Hessian can be simplified for large~$n$ yielding an
approximation to the log marginal likelihood of the form
\begin{equation}
\ln \Pr(x|\xmin) \simeq \Loglike - \half\xmin\ln n \punctspace ,
\end{equation}
where $\Loglike$ is the value of the conventional log-likelihood at its
maximum.  This type of approximation is known as a \defn{Bayesian
 information criterion} or BIC.  The maximum of the BIC with respect
to~$\xmin$ then gives the estimated value~$\hat{x}_{\min}$.\footnote{The
 same procedure of reducing the likelihood by $\half\ln n$ times the
 number of model parameters to avoid over-fitting can also be justified on
 non-Bayesian grounds for many model selection problems.}

This method works well under some circumstances, but can also present
difficulties.  In particular, the assumption that $\xmin-1$ parameters are
needed to model the data below $\xmin$ may be excessive: in many cases the
distribution below~$\xmin$, while not following a power law, can
nonetheless be represented well by a model with a much smaller number of
parameters.  In this case, the BIC tends to underestimate the value
of~$\xmin$ and this could result in biases on the subsequently calculated
value of the scaling parameter.  More importantly, it is also unclear how
the BIC (and similar methods) can be generalized to the case of continuous
data, for which there is no obvious choice for how many parameters are
needed to represent the empirical distribution below~$\xmin$.

Our second approach for estimating~$\xmin$, proposed by
Clauset~\etal~\cite{Clauset07}, can be applied to both discrete and
continuous data.  The fundamental idea behind this method is simple: we
choose the value of $\hat{x}_{\min}$ that makes the probability
distributions of the measured data and the best-fit power-law model as
similar as possible above~$\hat{x}_{\min}$.  In general, if we choose
$\hat{x}_{\min}$ higher than the true value~$\xmin$, then we are
effectively reducing the size of our data set, which will make the
probability distributions a poorer match because of statistical
fluctuation.  Conversely, if we choose $\hat{x}_{\min}$ smaller than the
true~$\xmin$, the distributions will differ because of the fundamental
difference between the data and model by which we are describing it.  In
between lies our best estimate.

There are a variety of measures for quantifying the distance between two
probability distributions, but for non-normal data the commonest is the
Kolmogorov-Smirnov or KS statistic~\cite{Numerical-Recipes-in-C}, which is
simply the maximum distance between the CDFs of the data and the fitted model:
\begin{equation}
 D = \max_{x\ge \xmin} \left| S(x) - P(x) \right| \punctspace.
\label{eq:ks}
\end{equation}
Here $S(x)$ is the CDF of the data for the observations with value at
least~$\xmin$, and $P(x)$ is the CDF for the power-law model that best fits
the data in the region $x\ge\xmin$.  Our estimate $\hat{x}_{\min}$~is then
the value of $\xmin$ that minimizes~$D$.\footnote{We note in passing that
 this approach can easily be generalized to the problem of estimating a
 lower cut-off for data following other (non-power-law) types of
 distributions.}


There is good reason to expect this method to produce reasonable results.
Note in particular that for right-skewed data of the kind we consider here
the method is especially sensitive to slight deviations of the data from
the power-law model around $\xmin$ because most of the data, and hence most
of the dynamic range of the CDF, lie in this region.  In practice, as we
show in the following section, the method appears to give excellent results
and generally performs better than the BIC approach.

\subsection{Tests of estimates for the lower bound}
\label{sec:lbtests}
\begin{figure}[t]
\begin{center}
\includegraphics[scale=0.4]{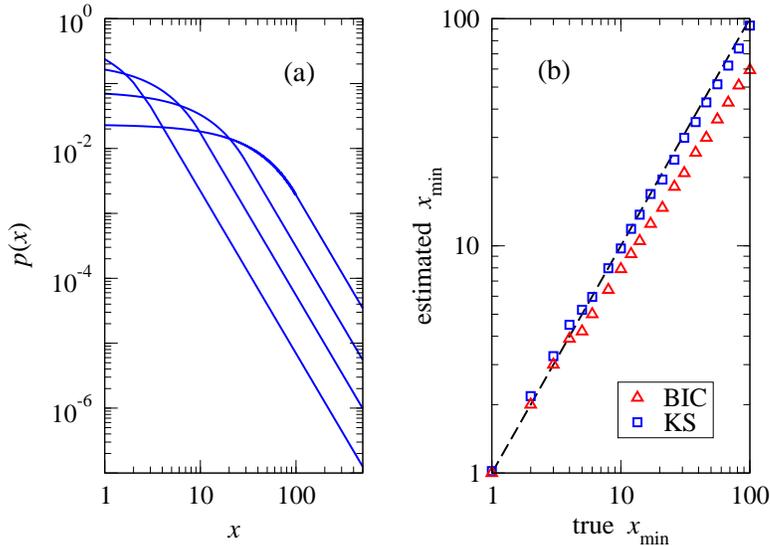}
\end{center}
\caption{(a)~Examples of the test distribution, Eq.~\eqref{eq:kstest}, used
 in the calculations described in the text, with power-law behavior for
 $x$ above $\xmin$ but non-power-law behavior below.  (b)~Value of
 $x_{\min}$ estimated using the Bayesian information criterion and KS
 approaches as described in the text, plotted as a function of the true
 value for discrete data with $n=50\,000$.  Results are similar for
 continuous data.}
\label{fig:xmin}
\end{figure}

As with our MLEs for the scaling parameter, we test our two methods for
estimating $\xmin$ by generating synthetic data and examining the methods'
ability to recover the known value of~$\xmin$.  For the tests presented
here we use synthetic data drawn from a distribution with the form
\begin{equation}
p(x) = \begin{cases}
      C (x/\xmin)^{-\alpha}     & \quad\text{for $x\ge \xmin$}\punctspace , \\
      C \e^{-\alpha(x/\xmin-1)} & \quad\text{for $x<\xmin$}\punctspace ,
      \end{cases}
\label{eq:kstest}
\end{equation}
with 
$\alpha=2.5$.  This distribution follows a power law at~$\xmin$ and above
but an exponential below.  Furthermore, it has a continuous slope at
$\xmin$ and thus deviates only gently from the power law as we pass below
this point, making for a challenging test.  Figure~\ref{fig:xmin}a shows a
family of curves from this distribution for different values of~$\xmin$.

In Fig.~\ref{fig:xmin}b we show the results of the application of both the
BIC and KS methods for estimating $\xmin$ to a large collection of data
sets drawn from Eq.~\eqref{eq:kstest}.  The plot shows the average
estimated value~$\hat{x}_{\min}$ as a function of the true~$\xmin$ for the
discrete case.  The KS method appears to give good estimates of $\xmin$ in
this case and performance is similar for continuous data also (not shown),
although the results tend to be slightly more conservative (i.e.,~to yield
slightly larger estimates~$\hat{x}_{\min}$).  The BIC method also performs
reasonably, but, as the figure shows, the method displays a tendency to
underestimate~$\xmin$, as we might expect given the arguments of the
previous section.  Based on these observations, we recommend the KS method
for estimating~$\xmin$ for general applications.

These tests used synthetic data sets of $n=50\,000$ observations, but good
estimates of $\xmin$ can be extracted from significantly smaller data sets
using the KS method; results are sensitive principally to the number of
observations in the power-law part of the distribution~$\ntail$.  For both
the continuous and discrete cases we find that good results can be achieved
provided we have about 1000 or more observations in this part of the
distribution.  This figure does depend on the particular form of the
non-power-law part of the distribution.  In the present test, the
distribution was designed specifically to make the determination of $\xmin$
challenging.  Had we chosen a form that makes a more pronounced departure
from the power law below $\xmin$ then the task of estimating
$\hat{x}_{\min}$ would be easier and presumably fewer observations would be
needed to achieve results of similar quality.


For some possible distributions there is, in a sense, no true value
of~$\xmin$.  The distribution $p(x) = C(x+k)^{-\alpha}$ follows a power law
in the limit of large~$x$, but there is no value of $\xmin$ above which it
follows a power law exactly.  Nonetheless, in cases such as this, we would
like our method to return an $\hat{x}_{\min}$ such that when we
subsequently calculate a best-fit value for $\alpha$ we get an accurate
estimate of the true scaling parameter.  In tests with such distributions
we find that the KS method yields estimates of~$\alpha$ that appear to be
asymptotically consistent, meaning that $\hat{\alpha}\to\alpha$ as
$n\to\infty$.  Thus again the method appears to work well, although it
remains an open question whether one can derive rigorous performance
guarantees.

Variations on the KS method are possible that use some other
goodness-of-fit measure that may perform better than the KS statistic under
certain circumstances.  The KS statistic is, for instance, known to be
relatively insensitive to differences between distributions at the extreme
limits of the range of~$x$ because in these limits the CDFs necessarily
tend to zero and one.  It can be reweighted to avoid this problem and be
uniformly sensitive across the range~\cite{Numerical-Recipes-in-C}; the
appropriate reweighting is
\begin{equation}
\label{eq:stabilized}
D^* = \max_{x\ge\hat{x}_{\min}}
     \frac{\left| S(x) - P(x) \right|}{\sqrt{P(x)(1-P(x))}} \punctspace .
\end{equation}
In addition a number of other goodness-of-fit statistics have been proposed
and are in common use, such as the Kuiper and Anderson-Darling
statistics~\cite{DAgostino-and-Stephens-on-GoF}.  We have performed tests
with each of these alternative statistics and find that results for the
reweighted KS and Kuiper statistics are very similar to those for the
standard KS statistic.  The Anderson-Darling statistic, on the other hand,
we find to be highly conservative in this application, giving estimates
$\hat{x}_{\min}$ that are too large by an order of magnitude or more.  When
there are many samples in the tail of the distribution this degree of
conservatism may be acceptable, but in most cases the reduction in the
number of tail observations greatly increases the statistical error on our
MLE for the scaling parameter and also reduces our ability to validate the
power-law model.

Finally, as with our estimate of the scaling parameter, we would like to
quantify the uncertainty in our estimate for~$\xmin$.  One way to do this
is to make use of a nonparametric ``bootstrap''
method~\cite{Efron-Tibshirani-bootstrap}.  Given our $n$ measurements, we
generate a synthetic data set with a similar distribution to the original
by drawing a new sequence of points $x_{i}$, $i=1\dots n$ uniformly at
random from the original data.  Using either method described above, we
then estimate $\xmin$ and $\alpha$ for this surrogate data set.  By taking
the standard deviation of these estimates over a large number of
repetitions of this process (say 1000), we can derive principled estimates
of our uncertainty in the original estimated parameters.

\subsection{Other techniques}
We would be remiss should we fail to mention some of the other techniques
in use for the analysis of power-law distributions, particularly those
developed within the statistics and finance communities, where the study of
these distributions has, perhaps, the longest history.  We give only a
brief summary of this material here; readers interested in pursuing the
topic further are encouraged to consult the books by
Adler~\etal~\cite{Adler-Feldman-Taqqu-heavy-tails} and
Resnick~\cite{Resnick06} for a more thorough explanation.\footnote{Another
 related area of study is ``extreme value theory,'' which concerns itself
 with the distribution of the largest or smallest values generated by
 probability distributions, values that assume some importance in studies
 of, for instance, earthquakes, other natural disasters, and the risks
 thereof---see~\cite{de-Hann-Ferreria-extreme-values}.}

In the statistical literature, researchers often consider a family of
distributions of the form
\begin{align}
\label{eq:heavy}
p(x) \propto L(x)\,x^{-\alpha} \punctspace,
\end{align}
where $L(x)$ is some slowly varying function, so that, in the limit of
large~$x$, $L(cx)/L(x)\to 1$ for any $c>0$.  An important issue in this
case---as in the calculations presented in this paper---is finding the
point~$\xmin$ at which the $x^{-\alpha}$ can be considered to dominate over
the non-asymptotic behavior of the function~$L(x)$, a task that can be
tricky if the data span only a limited dynamic range or if the
non-power-law behavior $|L(x)-L(\infty)|$ decays only a little faster
than~$x^{-\alpha}$.  In such cases, a visual approach---plotting an
estimate $\hat{\alpha}$ of the scaling parameter as a function of~$\xmin$
(called a Hill plot) and choosing for $\hat{x}_{\min}$ the value beyond
which $\hat{\alpha}$ appears stable---is a common technique.  Plotting
other statistics, however, can often yield better results---see, for
example, \cite{Kratz96} and~\cite{Stoev-Michailidis-Taqqu-max-self-sim}.
An alternative approach, quite common in the quantitative finance
literature, is simply to limit the analysis to the largest observed samples
only, such as the largest $\sqrt{n}$ or $\frac{1}{10}n$
observations~\cite{Farmer04}.

The methods described in Section~\ref{sec:xmin}, however, offer several
advantages over these techniques.  In particular, the KS method of
Section~\ref{sec:xmin} gives estimates of~$\xmin$ as least as good while
being simple to implement and having low enough computational costs that it
can be effectively used as a foundation for further analyses such as the
calculation of $p$-values in Section~\ref{sec:hypothesis}.  And, perhaps
more importantly, because the KS method removes the non-power-law portion
of the data entirely from the estimation of the scaling parameter, the fit
to the remaining data has a simple functional form that allows us to easily
test the level of agreement between the data and the best-fit model, as
discussed in Section~\ref{sec:alttesting}.

\section{Testing the power-law hypothesis}
\label{sec:hypothesis}

The tools described in the previous sections allow us to fit a power-law
distribution to a given data set and provide estimates of the
parameters~$\alpha$ and$~\xmin$.  They tell us nothing, however, about
whether the power law is a \emph{plausible} fit to the data.  Regardless of
the true distribution from which our data were drawn, we can always fit a
power law.  We need some way to tell whether the fit is a good match to the
data.

\begin{figure}[t]
\begin{center}
\includegraphics[scale=0.562]{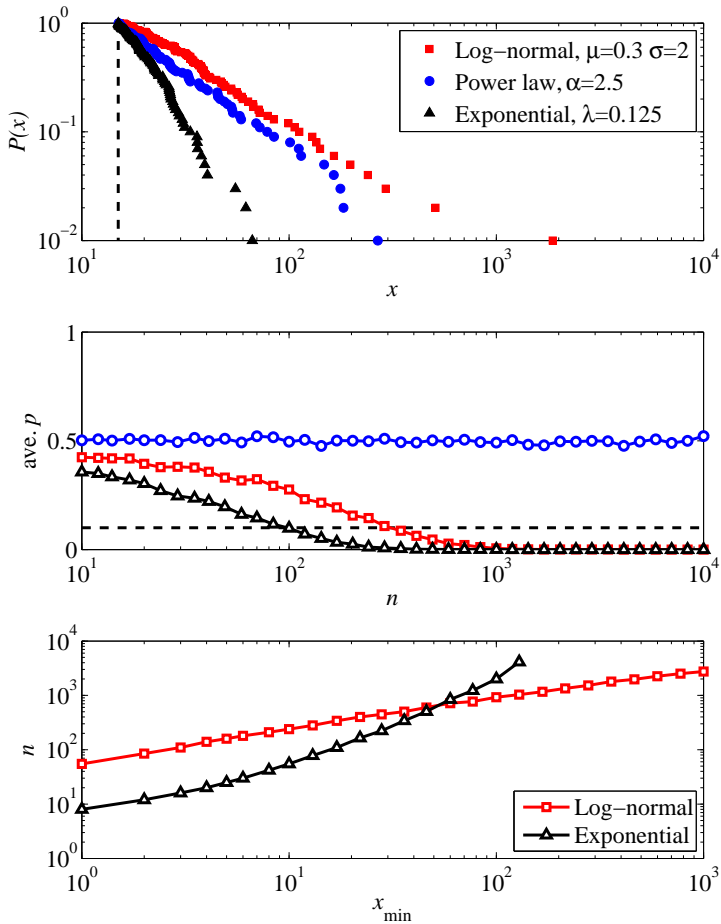}
\end{center}
\caption{(a)~The CDFs of three small samples ($n=100$) drawn from different
 continuous distributions: a log-normal with $\mu=0.3$ and $\sigma=2$, a
 power law with $\alpha=2.5$, and an exponential with $\lambda=0.125$, all
 with $\xmin=15$.  (Definitions of the parameters are as in
 Table~\ref{table:distributions}.)  Visually, each of the CDFs appears
 roughly straight on the logarithmic scales used, but only one is a true
 power law.  (b)~The average $p$-value for the maximum likelihood
 power-law model for samples from the same three distributions, as a
 function of the number of observations~$n$.  As $n$ increases, only the
 $p$-value for power-law distributed data remains above our rule-of-thumb
 threshold $p=0.1$ with the others falling off towards zero, indicating
 that $p$ does correctly identify the true power-law behavior in this
 case.  (c)~The average number of observations~$n$ required to reject the
 power-law hypothesis (i.e.,~to make $p<0.1$) for data drawn from the
 log-normal and exponential distributions, as a function of~$\xmin$.}
\label{fig:alts}
\end{figure}

Most previous empirical studies of ostensibly power-law distributed data
have not attempted to test the power-law hypothesis quantitatively.
Instead, they typically rely on qualitative appraisals of the data, based
for instance on visualizations.  But these can be deceptive and can lead to
claims of power-law behavior that do not hold up under closer scrutiny.
Consider Fig.~\ref{fig:alts}a, which shows the CDFs of three small data
sets ($n=100$) drawn from a power-law distribution with $\alpha=2.5$, a
log-normal distribution with $\mu=0.3$ and $\sigma=2.0$, and an exponential
distribution with exponential parameter $\lambda=0.125$.  In each case the
distributions have a lower bound of $\xmin=15$.  Because each of these
distributions looks roughly straight on the log-log plot used in the
figure, one might, upon cursory inspection, judge all three to follow power
laws, albeit with different scaling parameters.  This judgment would,
however, be wrong---being roughly straight on a log-log plot is a necessary
but not sufficient condition for power-law behavior.

Unfortunately, it is not straightforward to say with certainty whether a
particular data set has a power-law distribution.  Even if data are drawn
from a power law their observed distribution is extremely unlikely to
exactly follow the power-law form; there will always be some small
deviations because of the random nature of the sampling process.  The
challenge is to distinguish deviations of this type from those that arise
because the data are drawn from a non-power-law distribution.

The basic approach, as we describe in this section, is to sample many
synthetic data sets from a true power-law distribution, measure how far
they fluctuate from the power-law form, and compare the results with
similar measurements on the empirical data.  If the empirical data set is
much further from the power-law form than the typical synthetic one, then
the power law is not a plausible fit to the data.  Two notes of caution are
worth sounding.  First, the effectiveness of this approach depends on how
we measure the distance between distributions.  Here, we use the
Kolomogorov-Smirnov statistic, which typically gives good results, but in
principle another goodness-of-fit measure could be used in its place.
Second, it is of course always possible that a non-power-law process will,
as a result again of sampling fluctuations, happen to generate a data set
with a distribution close to a power law, in which case our test will fail.
The odds of this happening, however, dwindle with increasing~$n$, which is
the primary reason why one prefers large statistical samples when
attempting to verify hypotheses such as these.

\subsection{Goodness-of-fit tests}
\label{sec:testing}
Given an observed data set and a hypothesized power-law distribution from
which the data are drawn, we would like to know whether our hypothesis is a
plausible one, given the data.

A standard approach to answering this kind of question is to use a
\defn{goodness-of-fit test}, which generates a $p$-value that quantifies
the plausibility of the hypothesis.  Such tests are based on measurement of
the ``distance'' between the distribution of the empirical data and the
hypothesized model.  This distance is compared with distance measurements
for comparable synthetic data sets drawn from the same model, and the
$p$-value is defined to be the fraction of the synthetic distances that are
larger than the empirical distance.  If $p$ is large (close to~1), then the
difference between the empirical data and the model can be attributed to
statistical fluctuations alone; if it is small, the model is not a
plausible fit to the data.

As we have seen in Sections~\ref{sec:xmin} and~\ref{sec:lbtests} there are
a variety of measures for quantifying the distance between two
distributions.  In our calculations we use the Kolmogorov-Smirnov (KS)
statistic, which we encountered in Section~\ref{sec:xmin}.\footnote{One of
 the nice features of the KS statistic is that \emph{its} distribution is
 known for data sets truly drawn from any given distribution.  This allows
 one to write down an explicit expression in the limit of large $n$ for
 the $p$-value---see for example Ref.~\cite{Numerical-Recipes-in-C}.
 Unfortunately, this expression is only correct so long as the underlying
 distribution is fixed.  If, as in our case, the underlying distribution
 is itself determined by fitting to the data and hence varies from one
 data set to the next, we cannot use this approach, which is why we
 recommend the Monto Carlo procedure described here instead.}  In detail
our procedure is as follows.

First, we fit our empirical data to the power-law model using the methods
of Section~\ref{sec:fitting} and calculate the KS statistic for this fit.
Next, we generate a large number of power-law distributed synthetic data
sets with scaling parameter~$\alpha$ and lower bound~$\xmin$ equal to those
of the distribution that best fits the observed data.  We fit each
synthetic data set individually to its own power-law model and calculate
the KS statistic for each one relative to its own model.  Then we simply
count what fraction of the time the resulting statistic is larger than the
value for the empirical data.  This fraction is our $p$-value.

Note crucially that for each synthetic data set we compute the KS statistic
relative to the best-fit power law for that data set, not relative to the
original distribution from which the data set was drawn.  In this way we
ensure that we are performing for each synthetic data set the same
calculation that we performed for the real data set, a crucial requirement
if we wish to get an unbiased estimate of the $p$-value.

The generation of the synthetic data involves some subtleties.  To obtain
accurate estimates of $p$ we need synthetic data that have a distribution
similar to the empirical data below $\xmin$ but that follow the fitted
power law above~$\xmin$.  To generate such data we make use of a
semiparametric approach.  Suppose that our observed data set has
$n_\mathrm{tail}$ observations $x\ge\xmin$ and $n$ observations in total.
We generate a new data set with $n$ observations as follows.  With
probability $n_\mathrm{tail}/n$ we generate a random number~$x_i$ drawn
from a power law with scaling parameter~$\hat{\alpha}$ and $x\ge\xmin$.
Otherwise, with probability $1-n_\mathrm{tail}/n$, we select one element
uniformly at random from among the elements of the observed data set that
have $x<\xmin$ and set $x_i$ equal to that element.  Repeating the process
for all $i=1\ldots n$ we generate a complete synthetic data set that indeed
follows a power law above $\xmin$ but has the same (non-power-law)
distribution as the observed data below.

We also need to decide how many synthetic data sets to generate.  Based on
an analysis of the expected worst-case performance of the test, a good rule
of thumb turns out to be the following: if we wish our $p$-values to be
accurate to within about $\epsilon$ of the true value, then we should
generate at least $\frac14 \epsilon^{-2}$ synthetic data sets.  Thus, if we
wish our $p$-value to be accurate to about 2 decimal digits, we would
choose $\epsilon=0.01$, which implies we should generate about $2500$
synthetic sets.  For the example calculations described in
Section~\ref{sec:results} we used numbers of this order, ranging from 1000
to $10\,000$ depending on the particular application.

Once we have calculated our $p$-value, we need to make a decision about
whether it is small enough to rule out the power-law hypothesis or whether,
conversely, the hypothesis is a plausible one for the data in question.  In
our calculations we have made the relatively conservative choice that the
power law is ruled out if $p\le0.1$: that is, it is ruled out if there is a
probability of 1 in 10 or less that we would merely by chance get data that
agree as poorly with the model as the data we have.  (In other contexts,
many authors use the more lenient rule $p\le0.05$, but we feel this would
let through some candidate distributions that have only a very small chance
of really following a power law.  Of course, in practice, the particular
rule adopted must depend on the judgment of the investigator and the
circumstances at hand.\footnote{Some readers will be familiar with the use
 of $p$-values to confirm (rather than rule out) hypotheses for
 experimental data.  In the latter context, one quotes a $p$-value for a
 ``null'' model, a model \emph{other} than the model the experiment is
 attempting to verify.  Normally one then considers low values of~$p$ to
 be good, since they indicate that the null hypothesis is unlikely to be
 correct.  Here, by contrast, we use the $p$-value as a measure of the
 hypothesis we are trying to verify, and hence high values, not low, are
 ``good.''  For a general discussion of the interpretation of $p$-values,
 see~\cite{Mayo-Cox-frequentist}.})

It is important to appreciate that a large $p$-value does not necessarily
mean the power law is the correct distribution for the data.  There are (at
least) two reasons for this.  First, there may be other distributions that
match the data equally well or better over the range of $x$ observed.
Other tests are needed to rule out such alternatives, which we discuss in
Section~\ref{sec:alttesting}.

Second, as mentioned above, it is possible for small values of $n$ that the
empirical distribution will follow a power law closely, and hence that the
$p$-value will be large, even when the power law is the wrong model for the
data.  This is not a deficiency of the method; it reflects the fact that it is
genuinely harder to rule out the power law if we have very little data.
For this reason, high $p$-values should be treated with caution when $n$ is
small.

\subsection{Performance of the goodness-of-fit test}
\label{sec:ptest}

To demonstrate the utility of this approach, and to show that it can
correctly distinguish power-law from non-power-law behavior, we consider
data of the type shown in Fig.~\ref{fig:alts}a, drawn from continuous
power-law, log-normal, and exponential distributions.  In
Fig.~\ref{fig:alts}b we show the average $p$-value, calculated as above,
for data sets drawn from these three distributions, as a function of the
number of samples~$n$.  When $n$ is small, meaning $n\lesssim100$ in this
case, the $p$-values for all three distributions are above our threshold
of~0.1, meaning that the power-law hypothesis is not ruled out by our
test---for samples this small we cannot accurately distinguish the data
sets because there is simply not enough data to go on.  As the sizes of the
samples become larger, however, the $p$-values for the two non-power-law
distributions fall off and it becomes possible to say that the power-law
model is a poor fit for these data sets, while remaining a good fit for the
true power-law data set.

It is important to note, however, that, since we fit the power-law form to
only the part of the distribution above~$\xmin$, the value of~$\xmin$
effectively controls how many data points we have to work with.  If $\xmin$
is large then only a small fraction of the data set falls above it and thus
the larger the value of $\xmin$ the larger the total value of~$n$ needed to
reject the power law.  This phenomenon is depicted in Fig.~\ref{fig:alts}c,
which shows the value of $n$ needed to achieve the threshold value of
$p=0.1$ for the log-normal and exponential distributions, as a function
of~$\xmin$.

\section{Alternative distributions}
\label{sec:alttesting}
The method described in Section~\ref{sec:hypothesis} provides a reliable
way to test whether a given data set is plausibly drawn from a power-law
distribution.  However, the results of such tests don't tell the whole
story.  Even if our data are well fit by a power law it is still possible
that another distribution, such as an exponential or a log-normal, might
give a fit as good or better.  We can eliminate this possibility by using a
goodness-of-fit test again---we can simply calculate a $p$-value for a fit
to the competing distribution and compare it to the $p$-value for the power
law.

Suppose, for instance, that we believe our data might follow either a
power-law or an exponential distribution.  If we discover that the
$p$-value for the power law is reasonably large (say $p>0.1$) then the
power law is not ruled out.  To strengthen our case for the power law we
would like to rule out the competing exponential distribution, if possible.
To do this, we would find the best-fit exponential distribution, using the
equivalent for exponentials of the methods of Section~\ref{sec:fitting},
and the corresponding KS statistic, then repeat the calculation for a large
number of synthetic data sets and hence calculate a $p$-value.  If the
$p$-value is sufficiently small, we can rule out the exponential as a model
for our data.

By combining $p$-value calculations with respect to the power law and
several plausible competing distributions, we can in this way make a good
case for or against the power-law form for our data.  In particular, if the
$p$-value for the power law is high, while those for competing
distributions are small, then the competition is ruled out and, although we
cannot say absolutely that the power law is correct, the case in its favor
is strengthened.

We cannot of course compare the power-law fit of our data with fits to
every competing the distribution, of which there are an infinite number.
Indeed, as is usually the case with data fitting, it will almost always be
possible to find a class of distributions that fits the data better than
the power law if we define a family of curves with a sufficiently large
number of parameters.  Fitting the statistical distribution of data should
therefore be approached using a combination of statistical techniques like
those described here and prior knowledge about what constitutes a
reasonable model for the data.  Statistical tests can be used to rule out
specific hypotheses, but it is up to the researcher to decide what a
reasonable hypothesis is in the first place.

\subsection{Direct comparison of models}
\label{sec:LRT}
The methods of the previous section can tell us if either or both of two
candidate distributions---usually the power-law distribution and some
alternative---can be ruled out as a fit to our data or, if neither is ruled
out, which is the better fit.  In many practical situations, however, we
only want to know the latter---which distribution is the better fit.  This
is because we will normally have already performed a goodness-of-fit test
for the first distribution, the power law.  If that test fails and the
power law is rejected, then our work is done and we can move on to other
things.  If it passes, on the other hand, then our principal concern is whether
another distribution might provide a better fit.

In such cases, methods exist which can directly compare two distributions
against one another and which are considerably easier to implement than the
KS test.  In this section we describe one such method, the \defn{likelihood
 ratio test}.\footnote{The likelihood ratio test is not the only possible
 approach.  Others include fully Bayesian
 approaches~\cite{kass:raftery:1995}, cross-validation~\cite{stone:1974},
 or minimum description length (MDL)~\cite{grunwald:2007}.}

The basic idea behind the likelihood ratio test is to compute the
likelihood of the data under two competing distributions.  The one with the
higher likelihood is then the better fit.  Alternatively one can calculate
the ratio of the two likelihoods, or equivalently the
logarithm~$\mathcal{R}$ of the ratio, which is positive or negative
depending on which distribution is better, or zero in the event of a
tie.

The sign of the log likelihood ratio alone, however, will not definitively
indicate which model is the better fit because, like other quantities, it
is subject to statistical fluctuation.  If its true value, meaning its
expected value over many independent data sets drawn from the same
distribution, is close to zero, then the fluctuations could change the sign
of the ratio and hence the results of the test cannot be trusted.  In order
to make a firm choice between distributions we need a log likelihood ratio
that is sufficiently positive or negative that it could not plausibly be
the result of a chance fluctuation from a true result that is close to
zero.

To make a quantitative judgment about whether the observed value of
$\mathcal{R}$ is sufficiently far from zero, we need to know the size of
the expected fluctuations, i.e.,~we need to know the standard
deviation~$\sigma$ on~$\mathcal{R}$.  This we can estimate from our data
using a method proposed by
Vuong~\cite{Vuong-testing-non-nested-hypotheses}.  This method gives a
$p$-value that tells us whether the observed sign of $\mathcal{R}$ is
statistically significant.  If this $p$-value is small (say $p<0.1$) then
it is unlikely that the observed sign is a chance result of fluctuations
and the sign is a reliable indicator of which model is the better fit to
the data.  If $p$ is large on the other hand, the sign is not reliable and
the test does not favor either model over the other.  It is one of the
advantages of this approach that it can tell us not only which of two
hypotheses is favored, but also when the data are insufficient to favor
either of them.\footnote{In cases where we are unable to distinguish between
 two hypothesized distributions one could claim that there is really no
 difference between them: if both are good fits to the data then it makes
 no difference which one we use.  This may be true in some cases but it is
 certainly not true in general.  In particular, if we wish to extrapolate
 a fitted distribution far into its tail, to predict, for example, the
 frequencies of large but rare events like major earthquakes or meteor
 impacts, then conclusions based on different fitted forms can differ
 enormously even if the forms are indistinguishable in the domain covered
 by the actual data.  Thus the ability to say whether the data clearly
 favor one hypothesis over another can have substantial practical
 consequences.}  The simple goodness-of-fit test of the previous section
provides no equivalent indication when the data are
insufficient.\footnote{One alternative method for choosing between
 distributions, the Bayesian approach described
 in~\cite{Stouffer-Malmgren-Amaral-on-Barabasi}, is essentially equivalent
 to the likelihood ratio test, but without the $p$-value to tell us when
 the results are significant.  The Bayesian estimation used is equivalent
 to a smoothing, which to some extent buffers the results against the
 effects of fluctuations~\cite{CRS-bayes-as-evol}, but the method is not
 capable, itself, of saying whether the results could be due to
 chance~\cite{Mayo-error,Wasserman-in-Bayesian-Analysis}.}  The technical
details of the likelihood ratio test are described in
Appendix~\ref{appendix:lrt}.

\subsection{Nested hypotheses}
\label{sec:nested}
In some cases the distributions we wish to compare may be \defn{nested},
meaning that one family of distributions is a subset of the other.  The
power law and the power law with exponential cutoff in
Table~\ref{table:distributions} provide an example of such nested
distributions.  When distributions are nested it is always the case that
the larger family of distributions will provide a fit at least as good as
the smaller, since every member of the smaller family is also a member of
the larger.  In this case, a slightly modified likelihood ratio test is
needed to properly distinguish between such models, as described in
Appendix~\ref{appendix:lrt}.

\begin{figure}[t]
\begin{center}
\includegraphics[scale=0.575]{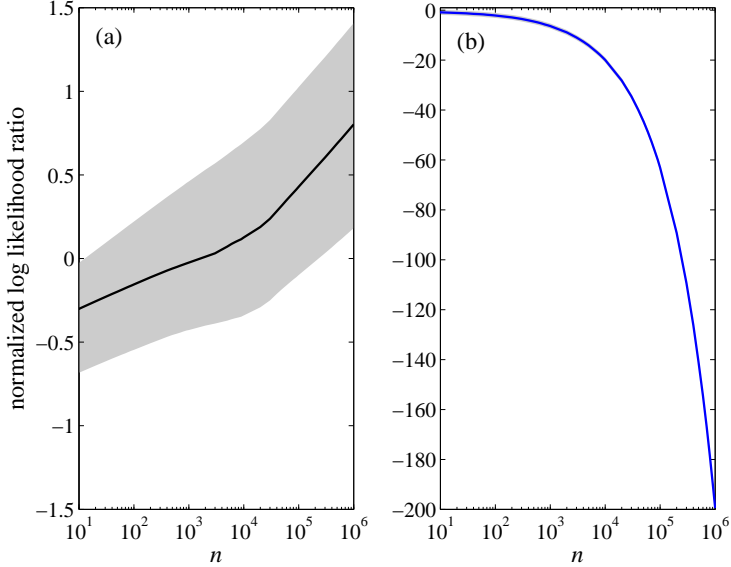}
\end{center}
\caption{Behavior of the normalized log likelihood ratio
 $n^{-1/2}\mathcal{R}/\sigma$ for synthetic data sets of $n$ points drawn
 from either (a)~a continuous power law with $\alpha=2.5$ and $\xmin=1$ or
 (b)~a log-normal with $\mu = 0.3$ and $\sigma = 2$.  Results are averaged
 over 1000 replications at each sample size, and the range covered by the
 $1^\mathrm{st}$ to $3^\mathrm{rd}$ quartiles is shown in gray.}
\label{fig:vuong-test-stat-behavior}
\end{figure}

\subsection{Performance of the likelihood-ratio test}
As with the other methods discussed here, we can quantify the performance
of the likelihood ratio test by applying it to synthetic data.  For our
tests, we generated data from two distributions, a continuous power law
with $\alpha=2.5$ and $\xmin=1$, and a log-normal distribution with
$\mu=0.3$ and $\sigma=2$ constrained to only produce positive values of~$x$
(These are the same parameter values we used in Section~\ref{sec:ptest}.)
In each case we drew $n$ independent values from each distribution and
estimated the value of~$\xmin$ for each set of values, then calculated the
likelihood ratio for the data above~$\xmin$ and the corresponding
$p$-value.  This procedure is repeated 1000 times to assess sampling
fluctuations.  Following Vuong~\cite{Vuong-testing-non-nested-hypotheses}
we calculate the normalized log likelihood ratio
$n^{-1/2}\mathcal{R}/\sigma$, where $\sigma$ is the estimated standard
deviation on~$\mathcal{R}$.  The normalized figure is in many ways more
convenient than the raw one since the $p$-value can be calculated directly
from it using Eq.~\eqref{eq:vuong}.  (In a sense this makes it unnecessary
to actually calculate~$p$ since the normalized log likelihood ratio
contains the same information, but it is convenient when making judgments
about particular cases to have the actual $p$-value at hand so we give both
in our results.)

Figure~\ref{fig:vuong-test-stat-behavior} shows the behavior of the
normalized log likelihood ratio as a function of~$n$.  As the figure shows,
it becomes increasing positive as~$n$ grows for data drawn from a true
power law, but increasingly negative for data drawn from a log-normal.


\begin{figure}[t]
\begin{center}
\includegraphics[scale=0.575]{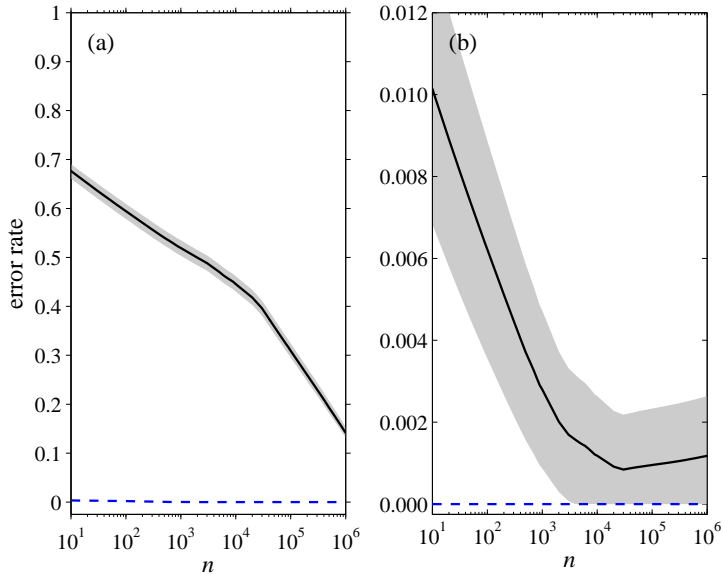}
\end{center}
\caption{Rates of misclassification of distributions by the likelihood
 ratio test if (a) the $p$-value is ignored and classification is based only
 on the sign of the log likelihood ratio, and (b) if the $p$-value is taken into account and we count only misclassifications where the log likelihood ratio has the wrong sign
 and the $p$-value is less than $0.05$.  Results are for the same
 synthetic data as Fig.~\ref{fig:vuong-test-stat-behavior}.  The black
 line shows the rate of misclassification (over 1000 repetitions) of power law samples as
 log-normals ($95\%$ confidence interval shown in grey), while the blue (dashed) line shows the rate of misclassification of
 log-normals as power laws ($95\%$ confidence interval is smaller than the width of the line).}
\label{fig:raw-error-rate-of-LRT}
\end{figure}

If we ignore the $p$-value and simply classify each of our synthetic data
sets as power-law or log-normal according to the raw sign of the log
likelihood ratio~$\mathcal{R}$ then, as we have said, we will sometimes
reach the wrong conclusion if $\mathcal{R}$ is close to zero and we are
unlucky with the sampling fluctuations.
Figure~\ref{fig:raw-error-rate-of-LRT}a shows the fraction of data sets
misclassified in this way in our tests as a function of~$n$ and though the
numbers decrease with sample size~$n$, they are uncomfortably large for
moderate values.  If we take the $p$-value into account, however, using its
value to perform a more nuanced classification as power-law, log-normal, or
undecided, as described above, the fraction of misclassifications is far
better, falling to a few parts per thousand, even for quite modest sample
sizes---see Fig.~\ref{fig:raw-error-rate-of-LRT}b.  These results indicate
that the $p$-value is effective at identifying cases in which the data are
insufficient to make a firm distinction between hypotheses.
\section{Applications to real-world data}
\label{sec:results}

In this section, as a demonstration of the utility of the methods described
in this paper, we apply them to a variety of real-world data sets
representing measurements of quantities whose distributions have been
conjectured to follow power laws.  As we will see, the results indicate
that some of the data sets are indeed consistent with a power-law
hypothesis, but others are not, and some are marginal cases for which the
power law is a possible candidate distribution, but is not strongly
supported by the data.

The twenty-four data sets we study are drawn from a broad variety of
different branches of human endeavor, including physics, earth sciences,
biology, ecology, paleontology, computer and information sciences,
engineering, and the social sciences.  They are as follows:
\begin{enumerate}
 \renewcommand{\theenumi}{\alph{enumi}}
 \renewcommand{\labelenumi}{\theenumi)}
\setlength{\itemsep}{0pt}
\item The frequency of occurrence of unique words in the novel \textit{Moby
   Dick} by Herman Melville~\cite{MEJN-on-power-laws}.
\item The degrees (i.e.,~numbers of distinct interaction partners) of
 proteins in the partially known protein-interaction network of the yeast
 \textit{Saccharomyces cerevisiae}~\cite{Ito00}.
\item The degrees of metabolites in the metabolic network of the bacterium
 \textit{Escherichia coli}~\cite{Holme06a}.
\item The degrees of nodes in the partially known network representation of
 the Internet at the level of autonomous systems for May
 2006~\cite{Holme06b}.  (An autonomous system is a group of IP addresses
 on the Internet among which routing is handled internally or
 ``autonomously,'' rather than using the Internet's large-scale border gateway protocol
 routing mechanism.)
\item The number of calls received by customers of AT\&T's long distance
 telephone service in the United States during a single day~\cite{ABW98,ACL00}.
\item The intensity of wars from 1816--1980 measured as the number of
 battle deaths per $10\,000$ of the combined populations of the warring
 nations~\cite{SS82,RT98}.
\item The severity of terrorist attacks worldwide from February 1968 to
 June 2006, measured as the number of deaths directly
 resulting~\cite{Clauset07}.
\item The number of bytes of data received as the result of individual web
 (HTTP) requests from computer users at a large research laboratory during
 a 24-hour period in June 1996~\cite{Willinger98}.  Roughly speaking this
 distribution represents the size distribution of web files transmitted
 over the Internet.
\item The number of species per genus of mammals.  This data set, compiled
 by Smith~\etal~\cite{Smith03}, is composed primarily of species alive
 today but also includes some recently extinct species, where ``recent''
 in this context means the last few tens of thousands of years.
\item The numbers of sightings of birds of different species in the North
 American Breeding Bird Survey for 2003.
\item The numbers of customers affected in electrical blackouts in the
 United States between 1984 and 2002~\cite{MEJN-on-power-laws}.
\item The numbers of copies of bestselling books sold in the United States
 during the period 1895 to 1965~\cite{Hackett67}.
\item The human populations of US cities in the 2000 US Census.
\item The sizes of email address books of computer users at a large
 university~\cite{NFB02}.
\item The sizes in acres of wildfires occurring on US federal land between
 1986 and 1996~\cite{MEJN-on-power-laws}.
\item Peak gamma-ray intensity of solar flares between 1980 and
 1989~\cite{MEJN-on-power-laws}.
\item The intensities of earthquakes occurring in California between 1910
 and 1992, measured as the maximum amplitude of motion during the
 quake~\cite{MEJN-on-power-laws}.
\item The numbers of adherents of religious denominations, bodies, and
 sects, as compiled and published on the web site {\tt adherents.com}.
\item The frequencies of occurrence of US family names in the 1990 US
 Census.
\item The aggregate net worth in US dollars of the richest individuals in
 the United States in October 2003~\cite{MEJN-on-power-laws}.
\item The number of citations received between publication and June 1997 by
 scientific papers published in 1981 and listed in the Science Citation
 Index~\cite{Redner-how-popular}.
\item The number of academic papers authored or coauthored by
 mathematicians listed in the American Mathematical Society's MathSciNet
 database.  (Data compiled by J. Grossman.)
\item The number of ``hits'' received by web sites from customers of the
 America Online Internet service in a single day~\cite{AH00b}.
\item The number of links to web sites found in a 1997 web crawl of about
 200 million web pages~\cite{Broder00}.
\end{enumerate}

Many of these data sets are only subsets of much larger entities (such as
the web sites, which are only a small fraction of the entire web).  In some
cases it is known that the sampling procedure used to obtain these subsets
may be biased, as, for example, in the protein
interactions~\cite{Sprinzak-Sattah-Margalit-reliability-of-protein-interaction},
citations and authorships~\cite{Bhattacharya-Getoor-entity-resolution}, and
the Internet~\cite{Achlioptas-et-al-bias-of-traceroute,caida07}.  We have
not attempted to correct any biases in our analysis.

\begin{figure}[!p]
\begin{center}
\includegraphics[scale=0.52]{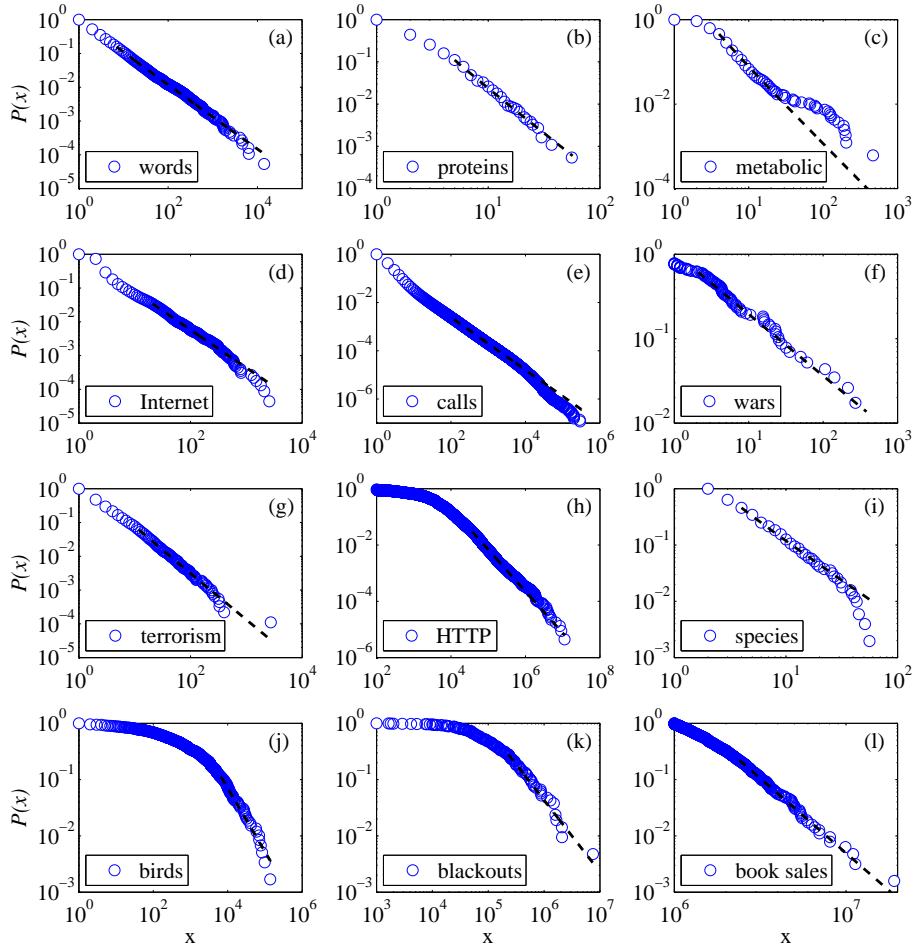}
\end{center}
\caption{The cumulative distribution functions $P(x)$ and their maximum
 likelihood power-law fits for the first twelve of our twenty-four
 empirical data sets.  (a)~The frequency of occurrence of unique words in
 the novel \textit{Moby Dick} by Herman Melville.  (b)~The degree
 distribution of proteins in the protein interaction network of the yeast
 \textit{S.~cerevisiae}.  (c)~The degree distribution of metabolites in
 the metabolic network of the bacterium \textit{E.~coli}.  (d)~The degree
 distribution of autonomous systems (groups of computers under single
 administrative control) on the Internet.  (e)~The number of calls
 received by US customers of the long-distance telephone carrier AT\&T.
 (f)~The intensity of wars from 1816--1980 measured as the number of
 battle deaths per $10\,000$ of the combined populations of the warring
 nations.  (g)~The severity of terrorist attacks worldwide from February
 1968 to June 2006, measured by number of deaths.  (h)~The number of bytes
 of data received in response to HTTP (web) requests from computers at a
 large research laboratory.  (i)~The number of species per genus of
 mammals during the late Quaternary period.  (j)~The frequency of
 sightings of bird species in the United States.  (k)~The number of
 customers affected by electrical blackouts in the United States.  (l)~The
 sales volume of bestselling books in the United States.}
\label{fig:data1}
\end{figure}

\begin{figure}[t]
\begin{center}
\includegraphics[scale=0.52]{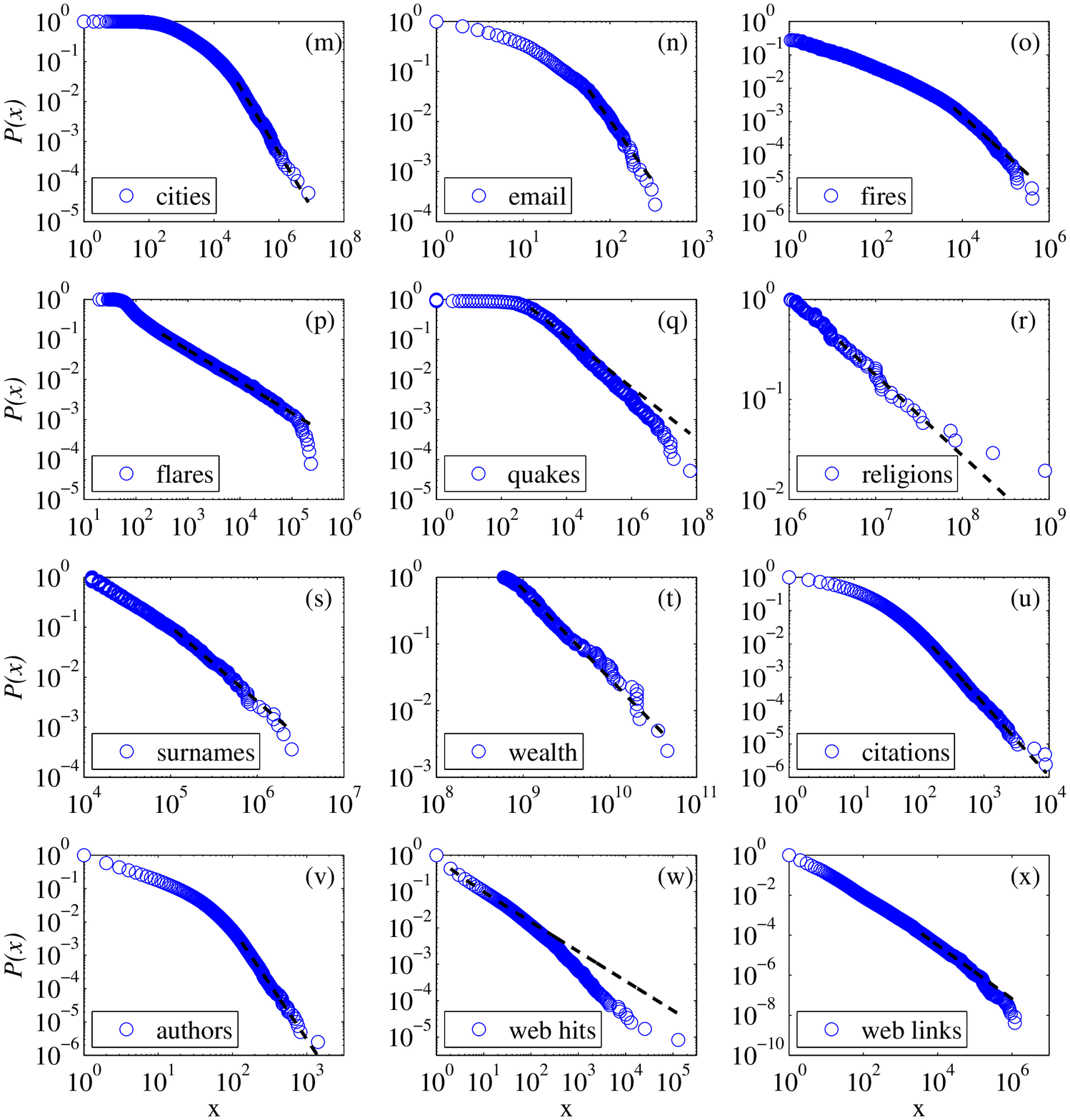}
\end{center}
\caption{The cumulative distribution functions $P(x)$ and their maximum
 likelihood power-law fits for the second twelve of our twenty-four
 empirical data sets.  (m)~The populations of cities in the United States.
 (n)~The sizes of email address books at a university.  (o)~The number of
 acres burned in California forest fires.  (p)~The intensities of solar
 flares.  (q)~The intensities of earthquakes.  (r)~The numbers of
 adherents of religious sects.  (s)~The frequencies of surnames in the
 United States.  (t)~The net worth in US dollars of the richest people in
 America.  (u)~The numbers of citations received by published academic
 papers.  (v)~The numbers of papers authored by mathematicians.  (w)~The
 numbers of hits on web sites from AOL users.  (x)~The numbers of
 hyperlinks to web sites.}
\label{fig:data2}
\end{figure}

In Table~\ref{table:data} we show results from the fitting of a power-law
form to each of these data sets using the methods described in
Section~\ref{sec:fitting}, along with a variety of generic statistics for
the data such as mean, standard deviation, and maximum value.  In the last
column of the table we give the $p$-value for the power-law model,
estimated as in Section~\ref{sec:hypothesis}, which gives a measure of how
plausible the power law is as a fit to the data.

As an indication of the importance of accurate methods for fitting
power-law data, we note that many of our values for the scaling parameters
differ considerably from those derived from the same data by previous
authors using \textit{ad hoc} methods.  For instance, the scaling parameter
for the protein interaction network of~\cite{Ito00} has been reported to
take a value of 2.44~\cite{YOB04}, which is quite different from, and
incompatible with, the value we find of $3.1\pm0.3$.  Similarly, the
citation distribution data of~\cite{Redner-how-popular} have been reported
to have a scaling parameter of either 2.9~\cite{TA99} or 2.5~\cite{KRL00},
neither of which are compatible with our maximum likelihood figure of
$3.16\pm0.06$.

The $p$-values in Table~\ref{table:data} indicate that 17 of the 24 data
sets are consistent with a power-law distribution.  The remaining seven
data sets all have $p$-values small enough that the power-law model can be
firmly ruled out.  In particular, the distributions for the HTTP
connections, earthquakes, web links, fires, wealth, web hits, and the
metabolic network cannot plausibly be considered to follow a power law; the
probability of getting by chance a fit as poor as the one observed is very
small in each of these cases and one would have to be unreasonably
optimistic to see power-law behavior in any of these data sets.  (For two
data sets---the HTTP connections and wealth distribution---the power law,
while not a good fit, is nonetheless better than the alternatives we tested
using the likelihood ratio test, implying that these data sets are not
well-characterized by any of the functional forms considered here.)

Tables~\ref{table:testing1} and~\ref{table:testing2} show the results of
likelihood ratio tests comparing the best fit power laws for each of our
data sets to the alternative distributions given in
Table~\ref{table:distributions}.  For reference, the first column repeats
the $p$-values given in Table~\ref{table:data}.  Based on the results of
our tests, we summarize in the final column of the table how convincing the
power-law model is as a fit to each data set.

\begin{sidewaystable}
\newcolumntype{e}{D{.}{.}{2}}
\newcolumntype{i}{D{.}{.}{0}}
\newcolumntype{p}{D{+}{\>\pm\>}{-1}}
\begin{center}
\begin{tabular}{li|eei|pep|c}
quantity & n & \langle x \rangle & \sigma & x_{\max} & \hat{x}_{\min} & \hat{\alpha} & n_{\rm tail} & $p$\\
\hline
count of word use		& 18\,855     & 11.14 & 148.33 & 14\,086 & 7+2 & 1.95(2) & 2958+987 & \tbf{0.49}\\ 
protein interaction degree 	& 1846        & 2.34 & 3.05 & 56 & 5+2 & 3.1(3) & 204+263  & \tbf{0.31}\\ 
metabolic degree		& 1641        & 5.68 & 17.81 & 468 & 4+1 & 2.8(1) & 748+136 & 0.00\\ 
Internet degree			& 22\,688     & 5.63 & 37.83 & 2583 & 21+9 & 2.12(9) & 770+1124 & \tbf{0.29}\\ 
telephone calls received	& 51\,360\,423& 3.88 & 179.09 & 375\,746 & 120+49 & 2.09(1) & \qquad102\,592+210\,147  & \tbf{0.63}\\ 
intensity of wars		& 115         & 15.70 & 49.97 & 382 & 2.1+3.5 & 1.7(2) & 70+14  & \tbf{0.20}\\ 
terrorist attack severity	& 9101        & 4.35 & 31.58 & 2749 & 12+4 & 2.4(2) & 547+1663  & \tbf{0.68} \\ 
HTTP size (kilobytes)		& 226\,386 & 7.36 & 57.94 & 10\,971 & 36.25+22.74 & 2.48(5) & 6794+2232  & 0.00\\ 
species per genus		& 509         & 5.59 & 6.94 & 56 & 4+2 & 2.4(2) & 233+138 & \tbf{0.10}  \\ 
bird species sightings		& 591         & 3384.36 & 10\,952.34 & 138\,705 & 6679+2463 & 2.1(2) & 66+41  & \tbf{0.55}\\ 
blackouts ($\times10^3$)	& 211         & 253.87 & 610.31 & 7500 & 230+90 & 2.3(3) & 59+35  & \tbf{0.62}\\ 
sales of books ($\times10^3$)	& 633         & 1986.67 & 1396.60 & 19\,077 & 2400+430 & 3.7(3) & 139+115 & \tbf{0.66} \\ 
\hline
population of cities ($\times10^3$) & 19\,447 & 9.00 & 77.83 & 8\,009 & 52.46+11.88 & 2.37(8) & 580+177   & \tbf{0.76}\\ 
email address books size	& 4581        & 12.45 & 21.49 & 333 & 57+21 & 3.5(6) & 196+449  & \tbf{0.16}\\ 
forest fire size (acres)	& 203\,785    & 0.90 & 20.99 & 4121 & 6324+3487 & 2.2(3) & 521+6801 & 0.05 \\ 
solar flare intensity		& 12\,773     & 689.41 & 6520.59 & 231\,300 & 323+89 & 1.79(2) & 1711+384  & \tbf{1.00} \\ 
quake intensity ($\times10^3$)	& 19\,302     & 24.54 & 563.83 & 63\,096 & 0.794+80.198 & 1.64(4) & 11\,697+2159  & 0.00\\ 
religious followers ($\times10^6$) & 103      & 27.36 & 136.64 & 1050 & 3.85+1.60 & 1.8(1) & 39+26  & \tbf{0.42}\\ 
freq.\ of surnames ($\times10^3$) & 2753      & 50.59 & 113.99 & 2502 & 111.92+40.67 & 2.5(2) & 239+215  & \tbf{0.20}\\ 
net worth (mil.\ USD)		& 400         & 2388.69 & 4\,167.35 & 46\,000 & 900+364 & 2.3(1) & 302+77  & 0.00 \\ 
citations to papers		& 415\,229    & 16.17 & 44.02 & 8904 & 160+35 & 3.16(6) & 3455+1859   & \tbf{0.20}\\ 
papers authored			& 401\,445    & 7.21 & 16.52 & 1416 & 133+13 & 4.3(1) & 988+377 & \tbf{0.90} \\ 
hits to web sites		& 119\,724    & 9.83 & 392.52 & 129\,641 & 2+13 & 1.81(8) & 50\,981+16\,898  & 0.00\\ 
links to web sites		& 241\,428\,853 & 9.15 & 106\,871.65 & 1\,199\,466 & 3684+151 & 2.336(9) & 28\,986+1560 & 0.00 \\ 
\end{tabular}
\end{center}
\caption{Basic parameters of the data sets described in this section, along
 with their power-law fits and the corresponding $p$-value (statistically
 significant values are denoted in \tbf{bold}).}
\label{table:data}
\end{sidewaystable}

\begin{sidewaystable}
\newcolumntype{e}{D{.}{.}{4}}
\setlength{\tabcolsep}{6pt}
\begin{center}
\begin{tabular}{l|c|cc|cc|cc|cc|l}
& power law & \multicolumn{2}{c|}{\text{log-normal}} & \multicolumn{2}{c|}{\text{exponential}} & \multicolumn{2}{c|}{\text{stretched exp.}} & \multicolumn{2}{c|}{\text{power law + cut-off}} & support for \\
data set & $p$ & \text{LR} & $p$ & \text{LR} & $p$ & \text{LR} & $p$ & \text{LR} & $p$ & power law \\
\hline
birds      & \tbf{0.55} & -0.850 &      0.40  & 1.87 & \tbf{0.06} & -0.882 &      0.38  & -1.24  &      0.12  & moderate \\
blackouts  & \tbf{0.62} & -0.412 &      0.68  & 1.21 &      0.23  & -0.417 &      0.68  & -0.382 &      0.38  & moderate \\
book sales & \tbf{0.66} & -0.267 &      0.79  & 2.70 & \tbf{0.01} & 3.885  & \tbf{0.00} & -0.140 &      0.60  & moderate \\
cities     & \tbf{0.76} & -0.090 &      0.93  & 3.65 & \tbf{0.00} & 0.204  &      0.84  & -0.123 &      0.62  & moderate \\
fires      &      0.05  & -1.78  & \tbf{0.08} & 4.00 & \tbf{0.00} & -1.82  & \tbf{0.07} & -5.02  & \tbf{0.00} & with cut-off \\
flares     & \tbf{1.00} & -0.803 &      0.42  & 13.7 & \tbf{0.00} & -0.546 &      0.59  & -4.52  & \tbf{0.00} & with cut-off \\
HTTP       &      0.00  &   1.77 & \tbf{0.08} & 11.8 & \tbf{0.00} & 2.65   & \tbf{0.01} & 0.000  &      1.00  & none \\
quakes     &      0.00  & -7.14  & \tbf{0.00} & 11.6 & \tbf{0.00} & -7.09  & \tbf{0.00} & -24.4  & \tbf{0.00} & with cut-off \\
religions  & \tbf{0.42} & -0.073 &      0.94  & 1.59 &      0.11  & 1.75   & \tbf{0.08} & -0.167 &      0.56  & moderate \\
surnames   & \tbf{0.20} & -0.836 &      0.40  & 2.89 & \tbf{0.00} & -0.844 &      0.40  & -1.36  & \tbf{0.10} & with cut-off \\
wars       & \tbf{0.20} & -0.737 &      0.46  & 3.68 & \tbf{0.00} & -0.767 &      0.44  & -0.847 &      0.19  & moderate \\
wealth     &      0.00  & 0.249  &      0.80  & 6.20 & \tbf{0.00} & 8.05   & \tbf{0.00} & -0.142 &      0.59  & none \\
web hits   &      0.00  & -10.21 & \tbf{0.00} & 8.55 & \tbf{0.00} & 10.94  & \tbf{0.00} & -74.66 & \tbf{0.00} & with cut-off \\
web links  &      0.00  & -2.24  & \tbf{0.03} & 25.3 & \tbf{0.00} & -1.08  &      0.28  & -21.2  & \tbf{0.00} & with cut-off
\end{tabular}
\end{center}
\caption{Tests of power-law behavior in the data sets with continuous
 (non-discrete) data.  (Results for the discrete data sets are given in
 Table~\ref{table:testing2}.)  For each data set we give a $p$-value for
 the fit to the power-law model and likelihood ratios for the
 alternatives.  We also quote $p$-values for the significance of each of
 the likelihood ratio tests.  Statistically significant $p$-values are
 denoted in \textbf{bold}.  Positive values of the log likelihood ratios
 indicate that the power-law model is favored over the alternative.  For
 non-nested alternatives, we give the normalized log likelihood ratio
 $n^{-1/2}\mathcal{R}/\sigma$ which appears in Eq.~\eqref{eq:vuong}, while
 for the power law with exponential cut-off we give the actual log
 likelihood ratio.  The final column of the table lists our judgment of
 the statistical support for the power-law hypothesis for each data set.
 ``None'' indicates data sets that are probably not power-law distributed;
 ``moderate'' indicates that the power law is a good fit but that there
 are other plausible alternatives as well; ``good'' indicates that the
 power law is a good fit and that none of the alternatives considered is
 plausible.  (None of the data sets in this table earned a rating of
 ``good,'' but one data set in Table~\ref{table:testing2}, for the
 frequencies of words, is so designated.)  In some cases we write ``with
 cut-off,'' meaning that the power law with exponential cutoff is clearly
 favored over the pure power law.  In each of the latter cases, however,
 some of the alternative distributions are also good fits, such as the
 log-normal or the stretched exponential distribution.}
\label{table:testing1}
\end{sidewaystable}

\begin{sidewaystable}
\setlength{\tabcolsep}{6pt}
\begin{center}
\begin{tabular}{l|c|cc|cc|cc|cc|cc|l}
& & \multicolumn{2}{c|}{Poisson} & \multicolumn{2}{c|}{log-normal} & \multicolumn{2}{c|}{exponential} & \multicolumn{2}{c|}{stretched exp.} & \multicolumn{2}{c|}{power law + cut-off} & support for\\
data set  & $p$ & LR & $p$ & LR & $p$ & LR & $p$ & LR & $p$ & LR & $p$ & power law\\
\hline
Internet  & \tbf{0.29} & $5.31$ & \tbf{0.00} & $-0.807$ &      0.42  & $6.49$  & \tbf{0.00} & $0.493$ &      0.62  & $-1.97$  & \tbf{0.05} & with cut-off \\ 
calls     & \tbf{0.63} & $17.9$ & \tbf{0.00} & $-2.03$  & \tbf{0.04} & $35.0$  & \tbf{0.00} & $14.3$  & \tbf{0.00} & $-30.2$  & \tbf{0.00} & with cut-off \\
citations & \tbf{0.20} & $6.54$ & \tbf{0.00} & $-0.141$ &      0.89  & $5.91$  & \tbf{0.00} & $1.72$  & \tbf{0.09} & $-0.007$ &      0.91  & moderate \\
email     & \tbf{0.16} & $4.65$ & \tbf{0.00} & $-1.10$  &      0.27  & $0.639$ &      0.52  & $-1.13$ &      0.26  & $-1.89$  & \tbf{0.05} & with cut-off \\
metabolic &      0.00  & $3.53$ & \tbf{0.00} & $-1.05$  &      0.29  & $5.59$  & \tbf{0.00} & $3.66$  & \tbf{0.00} & $0.000$  &      1.00  & none \\
papers    & \tbf{0.90} & $5.71$ & \tbf{0.00} & $-0.091$ &      0.93  & $3.08$  & \tbf{0.00} & $0.709$ &      0.48  & $-0.016$ &      0.86  & moderate \\ 
proteins  & \tbf{0.31} & $3.05$ & \tbf{0.00} & $-0.456$ &      0.65  & $2.21$  & \tbf{0.03} & $0.055$ &      0.96  & $-0.414$ &      0.36  & moderate \\
species   & \tbf{0.10} & $5.04$ & \tbf{0.00} & $-1.63$  &      0.10  & $2.39$  & \tbf{0.02} & $-1.59$ &      0.11  & $-3.80$  & \tbf{0.01} & with cut-off \\
terrorism & \tbf{0.68} & $1.81$ & \tbf{0.07} & $-0.278$ &      0.78  & $2.457$ & \tbf{0.01} & $0.772$ &      0.44  & $-0.077$ &      0.70  & moderate \\
words     & \tbf{0.49} & $4.43$ & \tbf{0.00} & $0.395$  &      0.69  & $9.09$  & \tbf{0.00} & $4.13$  & \tbf{0.00} & $-0.899$ &      0.18  & good
\end{tabular}
\end{center}
\caption{Tests of power-law behavior in the data sets with discrete
 (integer) data.  Statistically significant $p$-values are denoted in
 \textbf{bold}.  Results for the continuous data sets are given in
 Table~\ref{table:testing1}; see that table for a description of the
 individual column entries.}
\label{table:testing2}
\end{sidewaystable}

There is only one case---the distribution of the frequencies of occurrence
of words in English text---in which the power law appears to be truly
convincing, in the sense that it is an excellent fit to the data and none
of the alternatives carries any weight.

Among the remaining data sets we can rule out the exponential distribution
as a possible fit in all cases save three.  The three exceptions are the
blackouts, religions, and email address books, for which the power law is
favored over the exponential but the accompanying $p$-value is large enough
that the results cannot be trusted.  For the discrete data sets
(Table~\ref{table:testing2}) we can also rule out the Poisson distribution
in every case.

The results for the log-normal and stretched exponential distributions are
more ambiguous; in most cases the $p$-values for the log likelihood ratio
tests are sufficiently large that the results of the tests are
inconclusive.  In particular, the distributions for birds, books, cities,
religions, wars, citations, papers, proteins, and terrorism are plausible
power laws, but they are also plausible log-normals and stretched
exponentials.  In cases such as these, it is important to look at physical
motivating or theoretical factors to make a sensible judgment about the
which distributional form is more reasonable---we must consider whether
there is a mechanistic or other non-statistical argument favoring one
distribution or another.  The specific problem of the indistinguishability
of power laws and stretched exponentials has also been discussed by
Malevergne~\etal~~\cite{Malevergne-etal-stretched-exp}.

In some other cases the likelihood ratio tests do give conclusive answers.
For instance, the stretched exponential is ruled out for the book sales,
telephone calls, and citation counts, but is strongly favored over the
power law for the forest fires and earthquakes.  The log-normal, on the
other hand, is not ruled out for any of our data sets except the HTTP
connections.  In general, we find that it is extremely difficult to tell
the difference between log-normal and power-law behavior.  Indeed over
realistic ranges of $x$ the two distributions are very closely equal, so it
appears unlikely that any test would be able to tell them apart unless we
have an extremely large data set.  (See the results on synthetic data
reported in Section~\ref{sec:alttesting}.)

Finally, for almost a dozen data sets---the forest fires, solar flares,
earthquakes, web hits, web links, telephone calls, Internet, email address
books, and mammal species---the power-law with a cut-off is clearly favored
over the pure power law.  For surnames the cut-off form is also favored but
only weakly, as the $p$-value is very close to our threshold.  For the
remaining data sets, the large $p$-values indicate that there is no
statistical reason to prefer the cut-off form over the pure form.


\section{Conclusions}
The study of power laws spans many disciplines, including physics, biology,
engineering, computer science, the earth sciences, economics, political
science, sociology, and statistics.  Unfortunately, well founded methods
for analyzing power-law data have not yet taken root in all, or even most,
of these areas and in many cases hypothesized distributions are not tested
rigorously against the data.  This naturally leaves open the possibility
that apparent power-law behavior is, in some cases at least, the result of
wishful thinking.

In this paper we have argued that the common practice of identifying and
quantifying power-law distributions by the approximately straight-line
behavior of a histogram on a doubly logarithmic plot should not be trusted:
such straight-line behavior is a necessary but by no means sufficient
condition for true power-law behavior.  Instead we have presented a
statistically principled set of techniques that allow for the validation
and quantification of power laws.  Properly applied, these techniques can
provide objective evidence for or against the claim that a particular
distribution follows a power law.  In principle, they could also be
extended to other, non-power-law distributions as well, although we have
not given such an extension here.

We have applied the methods we describe to a large number of data sets from
various fields.  For many of these the power-law hypothesis turns out to
be, statistically speaking, a reasonable description of the data.  That is,
the data are compatible with the hypothesis that they are drawn from a
power-law distribution, although they are often compatible with other
distributions as well, such as log-normal or stretched exponential
distributions.  In the remaining cases the power-law hypothesis is found to
be incompatible with the observed data.  In some instances, such as the
distribution of earthquakes, the power law is plausible only if one assumes
an exponential cut-off that modifies the extreme tail of the distribution.

For some measured quantities, the answers to questions of scientific
interest may not rest upon the distribution following a power law
perfectly.  It may be enough, for example, that a quantity merely have a
heavy-tailed distribution.  In studies of the Internet, for instance, the
distributions of many quantities, such as file sizes, HTTP connections,
node degrees, and so forth, have heavy tails and appear visually to follow
a power law, but upon more careful analysis it proves impossible to make a
strong case for the power-law hypothesis; typically the power-law
distribution is not ruled out but competing distributions may offer a
better fit to the data.  Whether this constitutes a problem for the
researcher depends largely on his or her scientific goals.  For network
engineers, simply quantifying the heavy tail may be enough to allow them to
address questions concerning, for instance, future infrastructure needs or
the risk of overload from large but rare events.  Thus in some cases 
power-law behavior may not be fundamentally more interesting than any other
heavy-tailed distribution.  (In such cases, non-parametric estimates of the
distribution may be useful, though making such estimates for heavy-tailed
data presents special
difficulties~\cite{Markovitch-Krieger-nonparametric-heavy-tails}.)  If, on
the other hand, the goal is, say, to infer plausible mechanisms that might
underlie the formation and evolution of Internet structure or traffic
patterns, then it may matter greatly whether the observed quantity follows
a power law or some other form.

In closing, we echo comments made by Ijiri and
Simon~\cite{Ijiri-Simon-skew} more than thirty years ago and similar
thoughts expressed more recently by Mitzenmacher~\cite{Mitzenmacher06}.
They argue that the characterization of empirical distributions is only a
part of the challenge that faces us in explaining the causes and roles of
power laws in the sciences.  In addition we also need methods to validate
the models that have been proposed to explain those power laws.  They also
urge that, wherever possible, we consider to what practical purposes these
robust and interesting behaviors can be put.  We hope that the methods
given here will prove useful in all of these endeavors, and that these
long-held hopes will at last be fulfilled.

\section*{Acknowledgments}
The authors thank Sandra Chapman, Allen Downey, Doyne Farmer, Jake Hofman,
Luwen Huang, Kristina Klinkner, Joshua Ladau, Michael Mitzenmacher,
Cristopher Moore, Sidney Resnick, Stilian Stoev, Nick Watkins, Michael
Wheatland, Christopher Wiggins, and Maxwell Young for helpful conversations
and comments and Lada Adamic, Alison Boyer, Andrei Broder, Allen Downey,
Petter Holme, Mikael Huss, Joshua Karlin, Sidney Redner, Janet Wiener, and
Walter Willinger for kindly sharing data.  This work was supported in part
by the Santa Fe Institute (AC) and by grants from the James S. McDonnell
Foundation (CRS and MEJN) and the National Science Foundation (MEJN).


\begin{appendix}

\section{Linear regression and power laws}
\label{appendix:linear_regression}
The most common approach for testing empirical data against a hypothesized
power-law distribution is to observe that the power law $p(x) \sim
x^{-\alpha}$ implies the linear form
\begin{equation}
\log p(x) = \alpha\,\log x + c.
\end{equation}
The probability density $p(x)$ can be estimated by constructing a histogram
of the data (or alternatively one can construct the cumulative distribution
function by a simple rank ordering of the data) and the resulting function
can then be fitted to the linear form by least-squares linear regression.
The slope of the fit is interpreted as the estimate~$\hat{\alpha}$ of the
scaling parameter.  Many standard packages exist that can perform this kind
of fitting, provide estimates and standard errors for the slope, and
calculate the fraction~$r^2$ of variance accounted for by the fitted line,
which is taken as an indicator of the quality of the fit.

Although this procedure appears frequently in the literature there are
several problems with it.  As we saw in Section~\ref{sec:fitting}, the
estimates of the slope are subject to systematic and potentially large
errors (see Table~\ref{table:estimates} and Fig.~\ref{fig:bias}), but there
are a number of other serious problems as well.  First, errors are hard to
estimate because they are not well-described by the usual regression
formulas, which are based on assumptions that do not apply in this case.
For continuous data, this problem can be exacerbated by the choice of
binning scheme used to construct the histogram, which introduces an
additional set of free parameters.  Second, a fit to a power-law
distribution can account for a large fraction of the variance even when the
fitted data do not follow a power law, and hence high values of~$r^2$
cannot be taken as evidence in favor of the power-law form.  Third, the
fits extracted by regression methods usually do not satisfy basic
requirements on probability distributions, such as normalization, and hence
cannot be correct.

Let us look at each of these objections in a little more detail.

\subsection{Calculation of standard errors}
The ordinary formula for the calculation of the standard error on the slope
of a regression line is correct when the assumptions of linear regression
hold, which include independent, Gaussian noise in the dependent variable
at each value of the independent variable.  When fitting to the logarithm
of a histogram as in the analysis of power-law data, however, the noise,
though independent, is not Gaussian.  The noise in the frequency
estimates~$p(x)$ themselves is Gaussian (actually Poisson), but the noise
in their logarithms is not.  (For $\ln p(x)$ to have Gaussian fluctuations,
$p(x)$~would have to have log-normal fluctuations, which would violate the
central limit theorem.)  Thus the formula for the error is inapplicable in
this case.

For fits to the CDF the noise in the individual values $P(x)$ is Gaussian
(since it is the sum of independent Gaussian variables), but again the
noise in the logarithm is not.  Furthermore, the assumption of independence
now fails, because $P(x)=P(x+1)+p(x)$ and hence adjacent values of the CDF
are strongly correlated.  Fits to the CDF are, as we showed in
Section~\ref{sec:fitting}, empirically more accurate as a method for
determining the scaling parameter~$\alpha$, but this is not because the
assumptions of the fit are any more valid.  The improvement arises because
the statistical fluctuations in the CDF are typically much smaller than
those in the PDF.  The error on the scaling parameter is thus smaller but
this does not mean that the \emph{estimate} of the error is any better.
(In fact, it is typically a gross underestimate because of the failure to
account for the correlations.)

\subsection{Validation}
If our data are truly drawn from a power-law distribution and $n$ is large,
then the probability of getting a low $r^2$ in a straight-line fit is
small, so a low value of $r^2$ can be used to reject the power-law
hypothesis.  Unfortunately, as we saw in Section~\ref{sec:hypothesis},
distributions that are nothing like a power law can appear to follow a
power law for small samples and some, like the log-normal, can approximate
a power law closely over many orders of magnitude, resulting in high values
of~$r^2$.  And even when the fitted distribution approximates a power law
quite poorly, it can still account for a significant fraction of the
variance, although less than the true power law.  Thus, though a low $r^2$
is informative, in practice we rarely see a low~$r^2$, regardless of the
actual form of the distribution, so that the value of $r^2$ tells us
little.  In the terminology of statistical theory, the value of $r^2$ has
very little \defn{power} as a hypothesis test because the probability of
successfully detecting a violation of the power-law assumption is low.

\subsection{Regression lines are not valid distributions}
The CDF must take the value~1 at~$\xmin$ if the probability distribution
above $\xmin$ is properly normalized.  Ordinary linear regression, however,
does not incorporate such constraints and hence, in general, the regression
line does not respect them.  Similar considerations apply for the PDF,
which must integrate to~1 over the range from $\xmin$ to~$\infty$.
Standard methods exist to incorporate constraints like these into the
regression analysis~\cite{Weisberg-regression}, but they are not used to
any significant extent in the literature on power laws.

\section{Maximum likelihood estimators for the power law}
\label{appendix:pl_mle}
In this appendix we give derivations of the maximum likelihood estimators
for the scaling parameter of a power law.

\subsection{Continuous data}
In the case of continuous data the maximum likelihood estimator for the
scaling parameter, first derived (to our knowledge) by Muniruzzaman in
1957~\cite{Muniruzzaman-on-Pareto}, is equivalent to the well-known Hill
estimator~\cite{Hill75}.  Consider the continuous power-law distribution,
\begin{align}
\label{eq:cpl}
p(x) = \frac{\alpha-1}{\xmin}\left(\frac{x}{\xmin}  \right)^{-\alpha} ,
\end{align}
where $\alpha$ is the scaling parameter and $\xmin$ is the minimum value at
which power-law behavior holds.  Given a data set containing $n$
observations $x_i\geq \xmin$, we would like to know the value of $\alpha$
for the power-law model that is most likely to have generated our data.
The probability that the data were drawn from the model is proportional to
\begin{align}
 p(x\,|\,\alpha) = \prod_{i=1}^{n}
 \frac{\alpha-1}{\xmin}\left(\frac{x_{i}}{\xmin} \right)^{-\alpha} .
\end{align}
This probability is called the \emph{likelihood} of the data given the
model.  The data are most likely to have been generated by the model with
scaling parameter~$\alpha$ that maximizes this function.  Commonly we
actually work with the logarithm~$\mathcal{L}$ of the likelihood, which has
its maximum in the same place:
\begin{align}
\label{eq:loglike}
\Loglike &= \ln p(x\,|\,\alpha)
         = \ln \prod_{i=1}^{n} \frac{\alpha-1}{\xmin}
           \left(\frac{x_{i}}{\xmin}  \right)^{-\alpha} \nonumber\\
        &= \sum_{i=1}^{n} \left[ \ln(\alpha-1) - \ln \xmin
           - \alpha \ln \frac{x_{i}}{\xmin} \right] \nonumber\\
        &= n \ln(\alpha-1) - n \ln \xmin - \alpha
           \sum_{i=1}^{n}\ln \frac{x_{i}}{\xmin}.
\end{align}
Setting $\partial\Loglike/\partial\alpha=0$ and solving for $\alpha$, we
obtain the \emph{maximum likelihood estimate} or MLE for the scaling
parameter:
\begin{align}
\label{eq:cMLE}
\hat{\alpha} = 1 + n \Biggl[ \sum_{i=1}^n \ln {x_i\over \xmin}
              \Biggr]^{-1}.
\end{align}

\subsection{Formal results}
There are a number of formal results in mathematical statistics that
motivate and support the use of the MLE:
\medbreak
\begin{theorem}
 Under mild regularity conditions, if the data are independent,
 identically-distributed draws from a distribution with parameter~$\alpha$,
 then as the sample size $n\to\infty$, $\hat{\alpha}\to\alpha$ almost surely.
\label{theorem:consistency-of-mle}
\end{theorem}
\begin{proof}
See, for instance,~\cite{Pitman-basic-theory}.
\end{proof}

\medbreak
\begin{proposition}[\cite{Muniruzzaman-on-Pareto}]
 The maximum likelihood estimator $\hat{\alpha}$ of the continuous power
 law converges almost surely on the true $\alpha$.
 \label{prop:consistency-of-mle}
\end{proposition}
\begin{proof}
 It is easily verified that $\ln(x/\xmin)$ has an exponential distribution
 with rate $\alpha-1$.  By the strong law of large numbers, therefore,
 $\frac{1}{n}\sum_{i=1}^{n}{\ln{x_i \over \xmin}}$ converges almost surely
 on the expectation value of $\ln(x/\xmin)$, which is ${(\alpha -
   1)}^{-1}$.
\end{proof}

\medbreak
\begin{theorem}
 If the MLE is consistent, and there exists an interval $(\alpha -
 \epsilon, \alpha+\epsilon)$ around the true parameter value $\alpha$
 where, for any $\alpha_1, \alpha_2$ in that interval,
\begin{equation}
{\partial^3 \Loglike(\alpha_1)/\partial \alpha^3\over
 \partial^2 \Loglike(\alpha_2)/\partial \alpha^2}
\end{equation}
is bounded for all~$x$, then asymptotically $\hat{\alpha}$ has a Gaussian
distribution centered on $\alpha$, whose variance is $1/nI(\alpha)$, where
\begin{equation}
I(\alpha) = -\Expect{\frac{\partial^2 \log{p(X|\alpha)}}{\partial
   \alpha^2}},
\end{equation}
which is called the \emph{Fisher information} at~$\alpha$.  Moreover,
$\partial^2 \Loglike(\hat{\alpha})/\partial \alpha^2 \to I(\alpha)$.
\label{theorem:asymptotic-dist-of-mle}
\end{theorem}
\begin{proof}
 For the quoted version of this result,
 see~\cite[ch.~3]{Barndorff-Nielsen-and-Cox-inference-and-asymptotics}.
 The first version of a proof of the asymptotic Gaussian distribution of
 the MLE, and its relation to the Fisher information, may be found
 in~\cite{Fisher-introduces-maximum-likelihood}.
\end{proof}

\medbreak
\begin{proposition}[\cite{Muniruzzaman-on-Pareto}]
 The MLE of the continuous power law is asymptotically Gaussian, with
 variance ${(\alpha-1)}^2/n$.
\label{prop:asymp-dist-of-mle}
\end{proposition}
\begin{proof}
 Follows by application of the preceding theorem.  Simple calculation
 shows that $\partial^2 \log{\Loglike(\alpha)}/\partial \alpha^2 =
 -n{(\alpha - 1)}^{-2}$ and $\partial^3 \log{\Loglike(\alpha)}/\partial
 \alpha^3 = 2n{(\alpha - 1)}^{-3}$, so that the ratio in question is
 $2{(\alpha_2-1)}^2/{(\alpha_1 - 1)}^{3}$.  Since $\alpha > 1$, this ratio
 is bounded on any sufficiently small interval around any~$\alpha$ and the
 hypotheses of the theorem are satisfied.
\end{proof}

\medbreak
A further standard result, the \defn{Cram{\'e}r-Rao inequality}, asserts
that for \emph{any} unbiased estimator of~$\alpha$, the variance is at
least $1/nI(\alpha)$.  (See~\cite[\S 32.3]{Cramer}, or, for an elementary
proof,~\cite{Pitman-basic-theory}.)  The MLE is said to be
\defn{asymptotically efficient}, since it attains this lower bound.

Proposition~\ref{prop:asymp-dist-of-mle} yields approximate standard errors
and Gaussian confidence intervals for~$\hat{\alpha}$, becoming exact as $n$
becomes large.  Corrections depend on how $\xmin$ is estimated and on the
resulting coupling between that estimate and~$\hat{\alpha}$.  As the
corrections are~$\Ord(1/n)$, however, while the leading terms are
$\Ord(1/\sqrt{n})$, we have neglected them in the main text.  The
corrections can be deduced from the ``sampling distribution''
of~$\hat{\alpha}$, i.e.,~the distribution of deviations from $\alpha$ due
to finite-sample fluctuations.  (See~\cite{Cramer}
or~\cite{Wasserman-all-of-stats} for introductions to sampling
distributions.)  In general, the sampling distribution is hard to obtain
analytically, but it can be found by
bootstrapping~\cite{Wasserman-all-of-stats,Efron-Tibshirani-bootstrap}.  An
important exception is when $\xmin$ is either known \textit{a priori} or an
effective $\xmin$ is simply chosen by fiat (as in the Hill estimator).
Starting from the distribution of~$\ln{x}$, it is then easy to show that
$(\hat{\alpha}-1)/n$ has an inverse gamma distribution with shape parameter
$n$ and scale parameter $\alpha-1$.  This implies~\cite{JKB94} that
$\hat{\alpha}$ has a mean of $(n\alpha-1)/(n-1)$ and a standard deviation
of $n(\alpha-1)/(n-1)\sqrt{n-2}$, differing, as promised, from the
large-$n$ values by $\Ord(1/n)$.

\begin{figure}[t]
\begin{center}
\includegraphics[scale=0.45]{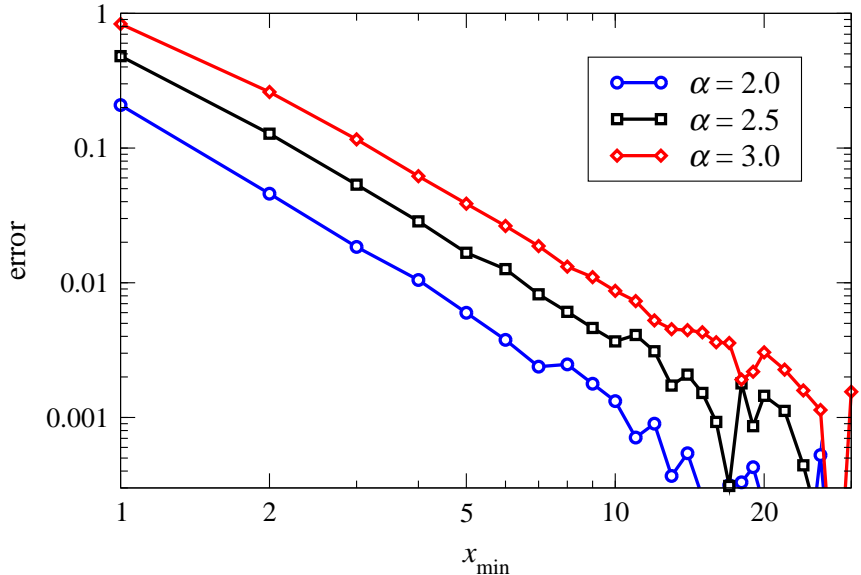}
\end{center}
\caption{The error on the estimated scaling parameter~$\hat{\alpha}$ that
 arises from using the approximate MLE for discrete data,
 Eq.~\eqref{eq:approxmle}, for $\alpha=2$, 2.5, and~3 (for 1000 repetitions), as a function
 of~$\xmin$.  The average error decays as $\Ord(\xmin^{-2})$ and becomes
 smaller than 1\% of the value of $\alpha$ when $\xmin\gtrsim6$.}
\label{fig:mlebias}
\end{figure}

\begin{figure}[t]
\begin{center}
\includegraphics[scale=0.45]{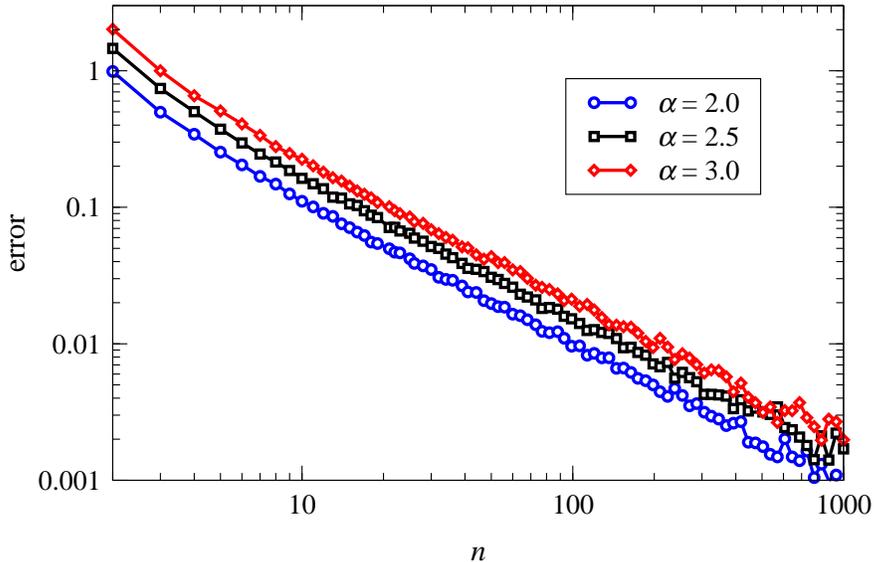}
\end{center}
\caption{The error on the estimated scaling parameter~$\hat{\alpha}$ from
 sample size effects for continuous data (similar results hold for the
 discrete case), for~$\alpha=2$, 2.5, and~3 (for 100 repetitions), as a function of sample size.
 The average error decays as $\Ord(n^{-1})$ and becomes smaller than 1\%
 of the value of $\alpha$ when $n\gtrsim50$.}
\label{fig:nconverge}
\end{figure}

\subsection{Discrete data}
We define the power-law distribution over an integer variable by
\begin{align}
p(x) = {x^{-\alpha}\over\zeta(\alpha,\xmin)},
\end{align}
where $\zeta(\alpha,\xmin)$ is the generalized or Hurwitz zeta function.
For the case $\xmin=1$, Seal~\cite{Seal-1952}~and, more recently,
Goldstein~\etal~\cite{Goldstein-Morris-Yen}~derived the maximum likelihood
estimator.  One can also derive an estimator for the more general case as
follows.

Following an argument similar to the one we gave for the continuous power law,
we can write down the log-likelihood function
\begin{equation}
\label{eq:dloglike}
\Loglike = \ln \prod_{i=1}^{n} \, {x_{i}^{-\alpha}\over
          \zeta(\alpha,\xmin)}
        = - n \ln \zeta(\alpha,\xmin) -
          \alpha\sum_{i=1}^{n} \ln x_{i}.
\end{equation}
Setting $\partial\Loglike/\partial\alpha = 0$ we then find
\begin{equation}
\frac{-n}{\zeta(\alpha,\xmin)}
\frac{\partial}{\partial\alpha}\zeta(\alpha,\xmin)  -
\sum_{i=1}^{n} \ln x_{i} = 0.
\end{equation}
Thus, the MLE~$\hat{\alpha}$ for the scaling parameter is the solution of
\begin{align}
\label{eq:dmle}
\frac{\zeta'(\hat{\alpha},\xmin)}{\zeta(\hat{\alpha},\xmin)} =
 - \frac{1}{n}\sum_{i=1}^{n} \ln x_{i} .
\end{align}
This equation can be solved numerically in a straightforward manner.
Alternatively, one can directly maximize the log-likelihood function
itself, Eq.~\eqref{eq:dloglike}.

The consistency and asymptotic efficiency of the MLE for the discrete power
law can be proved by applying Theorems \ref{theorem:consistency-of-mle} and
\ref{theorem:asymptotic-dist-of-mle}.  As the calculations involved are
long and messy, however, we omit them here.  Brave readers can
consult~\cite{Arnold-on-Pareto-distributions} for the details.

Equation~\eqref{eq:dmle} is somewhat cumbersome.  If $\xmin$ is moderately
large then a reasonable figure for $\alpha$ can be estimated using the much
more convenient approximate formula derived in the next section.

\subsection{Approximate estimator for the scaling parameter of the discrete
 power law} Given a differentiable function~$f(x)$, with indefinite
integral~$F(x)$, such that $F'(x)=f(x)$,
\begin{align}
& \int_{x-\frac12}^{x+\frac12} f(t) \>\d t
   = F\bigl(x+\half\bigr) - F\bigl(x-\half\bigr)
                                            \nonumber\\
 & \quad = \bigl[ F(x) + \half F'(x) + \mbox{$\frac18$} F''(x)
      + \mbox{$\frac{1}{48}$} F'''(x) \bigr] \nonumber\\
 & \qquad{} - \bigl[ F(x) - \half F'(x) + \mbox{$\frac18$} F''(x)
      - \mbox{$\frac{1}{48}$} F'''(x) \bigr] + \ldots \nonumber\\
 & \quad = f(x) + \mbox{$\frac{1}{24}$} f''(x) + \ldots
\end{align}
Summing over integer~$x$, we then get
\begin{align}
\int_{\xmin-\frac12}^\infty f(t) \>\d t &= \sum_{x=\xmin}^\infty f(x)
 + \frac{1}{24} \sum_{x=\xmin}^\infty f''(x) + \ldots
\end{align}
For instance, if $f(x)=x^{-\alpha}$ for some constant~$\alpha$, then we
have
\begin{align}
& \int_{\xmin-\frac12}^\infty t^{-\alpha} \>\d t
 = {\bigl(\xmin-\frac12\bigr)^{-\alpha+1}\over\alpha-1} \nonumber\\
&\quad = \sum_{x=\xmin}^\infty x^{-\alpha}
  + \frac{\alpha(\alpha+1)}{24} \sum_{x=\xmin}^\infty x^{-\alpha-2}
  + \ldots \nonumber\\
&\quad = \zeta(\alpha,\xmin)
  \bigl[ 1 + \Ord\bigl(\xmin^{-2}\bigr) \bigr],
\end{align}
where we have made use of the fact that $x^{-2}\le \xmin^{-2}$ for all
terms in the second sum.  Thus
\begin{equation}
\zeta(\alpha,\xmin) =
 {\bigl(\xmin-\frac12\bigr)^{-\alpha+1}\over\alpha-1}
 \bigl[ 1 + \Ord\bigl(\xmin^{-2}\bigr) \bigr].
\end{equation}
Differentiating this expression with respect to~$\alpha$, we also have
\begin{equation}
\zeta'(\alpha,\xmin) =
- {\bigl( \xmin-\half \bigr)^{-\alpha+1}\over\alpha-1}
\biggl[ {1\over\alpha-1} + \ln \bigl( \xmin - \half \bigr) \biggr]
\bigl[ 1 + \Ord\bigl(\xmin^{-2}\bigr) \bigr].
\end{equation}

We can use these expressions to derive an approximation to the maximum
likelihood estimator for the scaling parameter~$\alpha$ of the discrete
power law, Eq.~\eqref{eq:dmle}, valid when $x_{\min}$ is large.  The ratio
of zeta functions in Eq.~\eqref{eq:dmle} becomes
\begin{equation}
\frac{\zeta'(\hat{\alpha},\xmin)}{\zeta(\hat{\alpha},\xmin)} =
- \biggl[ {1\over\hat{\alpha}-1} + \ln \bigl( \xmin - \half \bigr) \biggr]
\bigl[ 1 + \Ord\bigl(\xmin^{-2}\bigr) \bigr],
\end{equation}
and, neglecting quantities of order $\xmin^{-2}$ by comparison with
quantities of order~1, we have
\begin{equation}
\hat{\alpha} \simeq 1 + n \Biggl[ \sum_{i=1}^n \ln {x_i\over \xmin-\half}
             \Biggr]^{-1},
\label{eq:approximate-discrete-MLE}
\end{equation}
which is in fact identical to the MLE for the continuous case except for
the $-\half$ in the denominator.

Numerical comparisons of Eq.~\eqref{eq:approximate-discrete-MLE} to the
exact discrete MLE, Eq.~\eqref{eq:dmle}, show
that Eq.~\eqref{eq:approximate-discrete-MLE} is a good approximation when
$\xmin\gtrsim6$---see Fig.~\ref{fig:mlebias}.

\section{Likelihood ratio tests}
\label{appendix:lrt}
Consider two different candidate distributions for our data set with
probability density functions $p_1(x)$ and~$p_2(x)$.  The likelihoods of
the data set within the two distributions are
\begin{equation}
\Like_1 = \prod_{i=1}^n p_1(x_i),\qquad
\Like_2 = \prod_{i=1}^n p_2(x_i),
\end{equation}
and the ratio of the likelihoods is
\begin{equation}
R = {\Like_1\over\Like_2} = \prod_{i=1}^n {p_1(x_i)\over p_2(x_i)}.
\end{equation}
Taking logs, the log likelihood ratio is
\begin{equation}
\mathcal{R} = \sum_{i=1}^n \bigl[ \ln p_1(x_i) - \ln p_2(x_i) \bigr]
           = \sum_{i=1}^n \bigl[ \ell_i^{(1)} - \ell_i^{(2)} \bigr],
\label{eq:likeratio}
\end{equation}
where $\ell_i^{(j)} = \ln p_j(x_i)$ can be thought of as the log-likelihood
for a single measurement~$x_i$ within distribution~$j$.

But since, by hypothesis, the~$x_i$ are independent, so also are the
differences $\ell_i^{(1)}-\ell_i^{(2)}$, and hence, by the central limit
theorem, their sum~$\mathcal{R}$ becomes normally distributed as $n$
becomes large, with expected variance $n\sigma^2$ where $\sigma^2$ is the
expected variance of a single term.  In practice we don't know the expected
variance of a single term, but we can approximate it in the usual way by
the variance of the data:
\begin{equation}
\sigma^2 = {1\over n} \sum_{i=1}^n
          \Bigl[ \bigl( \ell_i^{(1)} - \ell_i^{(2)} \bigr)
          - \bigl( \bar{\ell}^{(1)} - \bar{\ell}^{(2)} \bigr) \Bigr]^2,
\label{eq:ellsigma}
\end{equation}
with
\begin{equation}
\bar{\ell}^{(1)} = {1\over n} \sum_{i=1}^n \ell_i^{(1)},\qquad
\bar{\ell}^{(2)} = {1\over n} \sum_{i=1}^n \ell_i^{(2)}.
\end{equation}

Now suppose we are worried that the true expectation value of the log
likelihood ratio is in fact zero, so that the observed sign of~$\mathcal{R}$ is
a product purely of the fluctuations and cannot be trusted as an indicator of
which model is preferred.  The probability that the measured log likelihood
ratio has a magnitude as large or larger than the observed
value~$|\mathcal{R}|$ is given by
\begin{align}
p &= {1\over\sqrt{2\pi n\sigma^2}} \biggl[ \int_{-\infty}^{-|\mathcal{R}|}
    \e^{-t^2/2n\sigma^2} \>\d t + \int_{|\mathcal{R}|}^\infty
    \e^{-t^2/2n\sigma^2} \>\d t \biggr] \nonumber\\
 &= \bigl| \erfc (\mathcal{R}/\sqrt{2n}\sigma) \bigr|,
\label{eq:vuong}
\end{align}
where $\sigma$ is given by Eq.~\eqref{eq:ellsigma} and
\begin{equation}
\label{eq:erfc}
\erfc(z) = 1 - \erf(z) = {2\over\sqrt{\pi}} \int_z^\infty \e^{-t^2} \>\d t
\end{equation}
is the complementary Gaussian error function (a function widely available
in scientific computing libraries and numerical analysis programs).

This $p$-value gives us an estimate of the probability that we measured a
given value of~$\mathcal{R}$ when the true value of $\mathcal{R}$ is close
to zero (and hence is unreliable as a guide to which model is favored).  If
$p$ is small (say $p<0.1$) then our value for $\mathcal{R}$ is unlikely to
be a chance result and hence its sign can probably be trusted as an
indicator of which model is the better fit to the data.  (It does not
however mean that the model is a \emph{good} fit, only that it is better
than the alternative.)  If on the other hand $p$ is large, then the
likelihood ratio test is inadequate to discriminate between the
distributions in question.\footnote{Note that, if we are interested in
 confirming or denying the power-law hypothesis, then a small $p$-value is
 ``good'' in the likelihood ratio test---it tells us whether the test's
 results are trustworthy---whereas it is ``bad'' in the case of the KS
 test, where it tells us that our model is a poor fit to the data.}

The rigorous proof of these results involves some subtleties that we have
glossed over in our description.  In particular, the distributions that we
are dealing with are in our case fixed by fitting to the same data that are
the basis for the likelihood ratio test and this introduces correlations
between the data and the log-likelihoods that must be treated with care.
However, Vuong~\cite{Vuong-testing-non-nested-hypotheses} has shown that
the results above do hold even in this case, provided $p_1$ and $p_2$ come
from distinct, non-nested families of distributions and the estimation is
done by maximizing the likelihood within each family.  (There are also some
additional technical conditions on the models, but they hold for all the
models considered here.)

\subsection{Nested hypotheses}
\label{appendix:nested}
When the true distribution lies in the smaller family of distributions, the
best fits to both families converge to the true distribution as $n$ becomes
large.  This means that the individual differences
$\ell_i^{(1)}-\ell_i^{(2)}$ in Eq.~\eqref{eq:likeratio} each converge to
zero, as does their variance~$\sigma^2$.  Consequently the ratio
$|\mathcal{R}|/\sigma$ appearing in the expression for the $p$-value tends
to 0/0, and its distribution does not obey the simple central limit theorem
argument given above.  A more refined analysis, using a kind of
probabilistic version of L'Hopital's rule, shows that in fact $\mathcal{R}$
adopts a chi-squared distribution as $n$ becomes
large~\cite{Wilks-likelihood-ratio}.  One can use this result to calculate
a correct $p$-value giving the probability that the log likelihood ratio
takes the observed value or worse, if the true distribution falls in the
smaller family.  If this $p$-value is small, then the smaller family can be
ruled out.  If not, then the best we can say is that the there is no
evidence that the larger family is needed to fit to the data, although
neither can it be ruled out.  For a more detailed discussion of this
special case see, for instance, \cite{Vuong-testing-non-nested-hypotheses}.

\section{Generating power-law distributed random numbers}
\label{appendix:deviates}
It is often the case in statistical studies of probability distributions
that we wish to generate random numbers with a given distribution.  For
instance, in this paper we have used independent random numbers drawn from
power-law distributions to test how well our fitting procedures can
estimate parameters such as $\alpha$ and~$\xmin$.  How should we generate
such numbers?  There are a variety of possible methods, but perhaps the
simplest and most elegant is the \defn{transformation
 method}~\cite{Numerical-Recipes-in-C}.  The method can be applied to both
continuous and discrete distributions; we describe both variants in turn in
this section.

Suppose $p(x)$ is a continuous probability density from which we wish to
draw random reals~$x\ge\xmin$.  Typically we will have a source of random
reals~$r$ uniformly distributed in the interval $0\le r<1$, generated by
any of a large variety of standard pseudo-random number generators.  The
probability densities $p(x)$ and $p(r)$ are related by
\begin{equation}
p(x) = p(r) {\d r\over\d x} = {\d r\over\d x},
\end{equation}
where the second equality follows because $p(r)=1$ over the interval from
$0$ to~$1$.  Integrating both sides with respect to~$x$, we then get
\begin{equation}
\label{eq:transformation1}
P(x) = \int_x^\infty p(x') \>\d x' = \int_r^1 \d r' = 1 - r,
\end{equation}
or equivalently
\begin{equation}
\label{eq:transformation2}
x = P^{-1}(1-r),
\end{equation}
where $P^{-1}$ indicates the functional inverse of the cumulative
distribution function~$P$.  For the case of the power law, $P(x)$~is given
by Eq.~\eqref{eq:cdf:continuous} and we find that
\begin{equation}
\label{eq:gen:continuous}
x = \xmin(1-r)^{-1/(\alpha-1)},
\end{equation}
which can be implemented in straightforward fashion in most computer
languages.

The transformation method can also be used to generate random numbers from
many other distributions, though not all, since in some cases there is no
closed form for the functional inverse of the CDF.  Table~\ref{table:rng}
lists the equivalent of Eq.~\eqref{eq:gen:continuous} for a number of the
distributions considered in this paper.

For a discrete power law the equivalent of Eq.~\eqref{eq:transformation1}
is
\begin{equation}
\label{eq:gen:discrete}
P(x) = \sum_{x'=x}^\infty p(x') = 1 - r.
\end{equation}
Unfortunately, $P(x)$~is given by Eq.~\eqref{eq:cdf:discrete}, which cannot
be inverted in closed form, so we cannot write a direct expression
equivalent to Eq.~\eqref{eq:gen:continuous} for the discrete case.
Instead, we typically solve Eq.~\eqref{eq:gen:discrete} numerically by a
combination of ``doubling up'' and binary
search~\cite{Numerical-Recipes-in-C}.  That is, for a given random
number~$r$, we first bracket a solution~$x$ to the equation by the
following steps:
\begin{algorithmic}
 \STATE $x_2 \leftarrow \xmin$
 \REPEAT
   \STATE $x_1 \leftarrow x_2$
   \STATE $x_2 \leftarrow 2x_1$
 \UNTIL{$P(x_2)<1-r$}
\end{algorithmic}
\medbreak
\noindent where $\leftarrow$ indicates assignment.  In plain English, this
code snippet tests whether $r\in[x,2x)$, starting with $x=x_{\min}$ and
doubling repeatedly until the condition is met.  The end result is a range
of $x$ in which $r$ is known to fall.  We then pinpoint the solution within
that range by binary search.  We need only continue the binary search until
the value of~$x$ is narrowed down to $k\le x<k+1$ for some integer~$k$:
then we discard the noninteger part and the result is a power-law
distributed random integer.  The generalized zeta functions needed to
evaluate $P(x)$ from Eq.~\eqref{eq:cdf:discrete} are typically calculated
using special functions from standard scientific libraries.  These
functions can be slow, however, so for cases where speed is important, such
as cases where we wish to generate very many random numbers, it may be
worthwhile to store the first few thousand values of the zeta function in
an array ahead of time to avoid recalculating them frequently.  Only the
values for smaller~$x$ are worth precalculating in this fashion, however,
since those in the tail are needed only rarely.

\begin{table}
\setlength{\tabcolsep}{4pt}
\begin{center}
\begin{tabular}{l|c}
\multirow{2}{*}{name} & \multirow{2}{*}{random numbers} \\
                     &                                 \\
\hline\hline
   \multirow{2}{5em}{power law}
 & \multirow{2}{*}{$x=\xmin(1-r)^{-1/(\alpha-1)}$} \\
& \\
\hline
   \multirow{2}{5em}{exponential}
 & \multirow{2}{*}{$x = \xmin - {1\over\lambda} \ln (1-r)$} \\
& \\
\hline
   \multirow{2}{5em}{stretched exponential}
 & \multirow{2}{*}{$x = \Bigl[ \xmin^\beta - {1\over\lambda} \ln (1-r)
                    \Bigr]^{1/\beta}$} \\
& \\
\hline
   \multirow{2}{5em}{log-normal}
 & $\scriptstyle \rho=\sqrt{-2\sigma^2\ln(1-r_1)},$ $\scriptstyle \theta=2\pi r_2$ \\
 & $\scriptstyle x_1=\exp(\rho\sin\theta),$ $\scriptstyle x_2=\exp(\rho\cos\theta)$ \\
\hline
   \multirow{2}{5em}{power law with cutoff}
 & \multirow{2}{*}{see caption} \\
& \\
\end{tabular}
\end{center}
\caption{Formulas for generating random numbers~$x$ drawn from continuous
 distributions, given a source of uniform random numbers~$r$ in the range
 $0\le r<1$.  For the case of the log-normal, there is no simple
 closed-form expression for generating a \emph{single} random number,
 but the expressions given will generate two
 independent log-normally distributed random numbers~$x_1,x_2$, given two
 uniform numbers~$r_1,r_2$ as input.  For the case of the power law with
 cutoff there is also no closed-form expression, but one can generate an
 exponentially distributed random number using the formula above and then
 accept or reject it with probability $p$ or $1-p$ respectively, where
 $p=(x/\xmin)^{-\alpha}$.  Repeating the process until a number is accepted
 then gives an~$x$ with the appropriate distribution.}
\label{table:rng}
\end{table}

If great accuracy is not needed it is also possible, as in the previous
section, to approximate the discrete power law by a continuous one.  The
approximation has to be done in the right way, however, if we are to get
good results.  Specifically, to generate integers $x\ge \xmin$ with an
approximate power-law distribution, we first generate continuous power-law
distributed reals $y\ge \xmin-\half$ and then round off to the nearest
integer $x=\bigl\lfloor y+\half\bigr\rfloor$.  Employing
Eq.~\eqref{eq:gen:continuous}, this then gives
\begin{equation}
x = \Bigl\lfloor \bigl( \xmin - \half \bigr) 
   \bigl(1-r\bigr)^{-1/(\alpha-1)} + \half \Bigr\rfloor.
\label{eq:cdfcontapp}
\end{equation}
The approximation involved in this approach is largest for the smallest
value of~$x$, which is by definition~$\xmin$.  For this value the
difference between the true power-law distribution,
Eq.~\eqref{eq:full:discrete}, and the approximation is given by
\begin{equation}
\Delta p = 1 - \Biggl( {\xmin+\frac12\over \xmin-\frac12}
              \Biggr)^{-\alpha+1} -
          {\xmin\over\zeta(\alpha,\xmin)}.
\end{equation}
For instance, when $\alpha=2.5$, this difference corresponds to an error of
more than 8\% on the probability $p(x)$ for $\xmin=1$, but the error
diminishes quickly to less than 1\% for $\xmin=5$, and less than 0.2\% for
$\xmin=10$.  Thus the approximation is in practice a reasonably good one
for quite modest values of~$\xmin$.  (Almost all of the data sets
considered in Section~\ref{sec:results}, for example, have $\xmin>5$.)  For
very small values of $\xmin$ the true discrete generator should still be
used unless large errors can be tolerated.  Other approximate approaches
for generating integers, such as rounding down (truncating) the value
of~$y$, give substantially poorer results and should not be used.

\begin{table}
\begin{center}
\setlength{\tabcolsep}{4pt}
\begin{tabular}{l|cc|ccc}
   & \multicolumn{2}{c|}{continuous} & \multicolumn{3}{c}{discrete} \\
$x$ & theory & generated & theory & generated & approx. \\
\hline
5   & 1.000  & 1.000     & 1.000  & 1.000     & 1.000 \\
6   & 0.761  & 0.761     & 0.742  & 0.740     & 0.738 \\
7   & 0.604  & 0.603     & 0.578  & 0.578     & 0.573 \\
8   & 0.494  & 0.493     & 0.467  & 0.466     & 0.463 \\
9   & 0.414  & 0.413     & 0.387  & 0.385     & 0.384 \\
10  & 0.354  & 0.352     & 0.328  & 0.325     & 0.325 \\
15  & 0.192  & 0.192     & 0.174  & 0.172     & 0.173 \\
20  & 0.125  & 0.124     & 0.112  & 0.110     & 0.110 \\
50  & 0.032  & 0.032     & 0.028  & 0.027     & 0.027 \\
100 & 0.011  & 0.011     & 0.010  & 0.010     & 0.009 
\end{tabular}
\end{center}
\caption{CDFs of discrete and continuous power-law distributions with
 $\xmin=5$ and $\alpha=2.5$.  The second and fourth columns show the
 theoretical values of the CDFs for the two distributions, while the third
 and fifth columns show the CDFs for sets of $100\,000$ random numbers
 generated from the same distributions using the transformation technique
 described in the text.  The final column shows the CDF for $100\,000$
 numbers generated using the continuous approximation to the discrete
 distribution, Eq.~\eqref{eq:cdfcontapp}.}
\label{table:deviates}
\end{table}

As an example of these techniques, consider continuous and discrete power
laws having $\alpha=2.5$ and $\xmin=5$.  Table~\ref{table:deviates} gives
the cumulative density functions for these two distributions, evaluated at
integer values of~$x$, along with the corresponding cumulative density
functions for three sets of $100\,000$ random numbers generated using the
methods described here.  As the table shows, the agreement between the
exact and generated CDFs is good in each case, although there are small
differences because of statistical fluctuations.  For numbers generated
using the continuous approximation to the discrete distribution the errors
are somewhat larger than for the exact generators, but still small enough
for many practical applications.

\end{appendix}


\end{document}